# Towards Real-time Structural Dynamics Simulation with Graph-based Digital Twin Modelling

Jun Zhang[1,#], Tong Zhang[2,#], Ying Wang[*1,3]

[1]School of Intelligent Civil and Ocean Engineering, Harbin Institute of Technology, Shenzhen, Guangdong, China

[2]Peng Cheng Laboratory, Shenzhen, Guangdong, China

[3]Guangdong Provincial Key Laboratory of Intelligent and Resilient Structures for Civil Engineering, Shenzhen, Guangdong, China

[#]Jun Zhang and Tong Zhang contributed equally to this work.

[*]*Correspondence Email*: yingwang@hit.edu.cn

## Abstract

Precise and timely simulation of a structure's dynamic behavior is crucial for evaluating its performance and assessing its health status. Traditional numerical methods are often limited by high computational costs and low efficiency, while deep learning approaches offer a promising alternative. However, these data-driven methods still face challenges, such as limited physical interpretability and difficulty in adapting to diverse structural configurations. To address these issues, this study proposes a graph-based digital twin modelling (GDTM) framework to simulate structural dynamic responses across various spatial topologies. In this framework, the adjacency matrix explicitly represents the spatial relationships between structural vertices, enhancing the model's physical interpretability. The effectiveness of the proposed framework was validated through comprehensive numerical and experimental studies. The results demonstrate that the framework accurately simulated structural dynamics across different topological configurations, with Normalized Mean-Squared Error (NMSE) values consistently below 0.005 in numerical simulations and 0.0015 in experimental validations. Furthermore, the framework achieved over 80-fold improvements in computational efficiency compared to traditional finite element methods (FEM). This research promotes the practical application of graph-based structural dynamics modelling, which has the potential to significantly advance structural performance evaluation and health monitoring.

# 1. Introduction

Civil infrastructure plays a fundamental and supportive role in the socio-economic development. Structural health conditions and structural performance under various loads, particularly dynamic ones, are critical for civil infrastructure, which determines the long-term reliability throughout its service life. To assess structural performance and health conditions, traditional methods evaluate whether the internal forces or deformations exceed the design values, which are normally expressed as a safety index[1-3]. Recently, increasing attention has been given to structural damage/condition. Examples include statistical models of time-domain signals[4], pattern recognition, or classification methods. These approaches typically utilize vibration signals to effectively evaluate the overall or local condition of structures. Their performance is constrained by the quality of the signals and the availability of measurement points. Wang and Hao[5] pointed out that using compressed sensing technology can significantly enhance the quality of sampled signals[6,7] and improve the accuracy of pattern recognition[8]. Additionally, the powerful feature extraction capability of deep learning models enables high-accuracy detection of structural damage[9,10], and even to some extent, enables the application of damage detection models across multiple data domains[11,12]. Nevertheless, the aforementioned methods may lack the capability to provide detailed insights for refined structural condition identification and health assessment. Therefore, numerical models capable of accurately, efficiently, and adaptively capturing the behavior of both entire structures and individual components—commonly referred to as digital twins (DT)—are indispensable.

The construction of numerical models or DT models normally utilize known physical laws and theoretical formulas. For example, the Discrete Element Method is commonly used to simulate the mechanical behavior of granular materials[13,14], such as sand, gravel, and rock, as well as fluid flow, and has been applied to simulate structural failure processes[15,16]. The Boundary Element Method requires discretization only on the boundary of the structure, making it particularly advantageous for simulating problems with infinite domains, such as foundations, tunnels, and pile foundations[17,18]. The Spectral Element Method, renowned for its computational accuracy and efficiency, is particularly advantageous for solving wave propagation problems[19-22] and vibration analysis[23,24], especially in scenarios where high precision and reduced computational costs are essential. Although these methods have been widely used, they are normally advantageous in specific applications only. As a general-purpose modelling tool, the FEM is more versatile and remains the dominant numerical technique for modeling the behavior of solid structures[25-27]. It uses advanced numerical solutions to solve differential equations with complex geometries, boundary conditions and loads. Moreover, its flexibility in handling various material properties, nonlinearities, and coupled physics problems makes it a powerful tool for simulating real-world engineering systems. Nevertheless, FEM, as well as aforementioned methods, face substantial computational challenges, especially in terms of time and memory requirements, when dealing with refined numerical models with a large amount of elements and/or timesteps. More importantly, these methods may be inaccurate when the system physics are unknown or too complex to capture, i.e., when the parameters needed to accurately describe the system are difficult to obtain.

To tackle these challenges, reduced order modeling and model updating techniques have been extensively studied. The former aims to generate simplified system representations that preserve main dynamic characteristics while minimizing the loss of critical information[28-30]. They not only significantly reduce computational costs but also enable the exploration of state space representations for complex systems[31,32]. The latter aims to accurately simulate the dynamic responses of structures[33,34] by adapting the numerical model through an optimization process which minimize the discrepancy between the model and the monitoring/test data.

However, in such a context, existing methods cannot fulfill the requirement of a DT model for accurately evaluating structural states due to their low accuracy, limited efficiency, and poor adaptability. Data-driven methods are promising due to their ability to efficiently model physics, including structural behavior, based on sampled data from either measurements or simulations. Traditional reduced order modeling methods, such as Principal Component Analysis[35,36], Proper Orthogonal Decomposition[37], and support vector machine[38,39], are data-driven methods and have been used in structural dynamic simulations. Based on the universal approximation theorem, neural networks can approximate any function[40]. Therefore, neural networks or deep learning (DL) algorithms, the most powerful machine learning technique, have been employed to simulate and analyze the dynamic behavior of structures. For example, Multilayer perceptrons (MLP) [41,42] and convolutional neural networks[43,44] have been used to predict/simulate both linear and nonlinear dynamic responses of structures. Recurrent neural networks and long short-term memory networks[45-47], known for their effectiveness in modeling sequential data by constructing time-stepped loops, have been used to predict/simulate dynamic structural responses.

Nevertheless, data-driven methods often lacked physical interpretability and struggled to handle diverse structural configurations. To address these challenges, this study proposed the GDTM approach for real-time structural dynamics simulation. The method processed graph-structured data, efficiently capturing complex patterns through diverse configurations of vertices and their connections. Moreover, the proposed method possessed clear physical significance, indicating its promising applications in structural performance evaluation and structural health monitoring. This paper is organized as follows: Section 2 introduces the related work. Section 3 describes the methodology, and Section 4 uses numerical and experimental models to validate the effectiveness of the method. Finally, Finally, the current work is concluded in Section 5.

## 2. Related works

The objective of this study is to address the limitations of existing methods in developing digital twins for simulating structural dynamic responses. To achieve this, a data-driven digital twin modelling approach is proposed for real-time structural dynamics simulation. Graph Neural Networks (GNNs) have been extensively studied across various domains, especially in representing topologies under complex layout conditions[48]. As an effective tool for analyzing graph-structured data, GNNs have been applied in areas such as social networks, chemical compounds, and the Internet of Things, and they show great potential to achieve superior physical interpretability and generalization capability compared to traditional machine learning and deep learning algorithms.

Basic GNN can roughly be categorized into two types: GCN[49], which updates features using fixed neighbor connections, making it suitable for stable, static graph structures; and GAT[50], which employs a dynamic adjacency matrix and uses attention mechanisms to adaptively assign weights to neighbors, offering greater flexibility for heterogeneous and complex graph structures. GraphSAGE[51], the variant of GCN, generated graph embeddings by sampling neighboring vertices, making them particularly well-suited for large-scale and evolving graph environments.

In the field of digital twin modelling, an interaction network, based on GNN principles, was used to model the physical trajectories of rigid bodies[52]. This work introduced a novel framework for learning physical interactions, demonstrating strong performance in various physics-based tasks. However, it faced challenges related to long-term prediction accuracy due to error accumulation and high computational complexity for large systems. Likewise, a Graph Network-based Simulator simulated interactions among fluids, rigid solids, and deformable materials[53]. This framework achieved excellent performance in modeling complex physics across multiple domains, though it heavily relied on particle representations and encountered scalability issues with increasing connection radii. Moreover, Zhang et al.[54] developed a GNN framework specifically for structures with varying topologies, effectively capturing dynamic responses but exhibiting limited generalizability to non-modular structures. Song et al. [55] proposed a physics-driven GNN model for elastic structural analysis without requiring labeled data, enhancing computational efficiency while being restricted to elastic systems. The GraphSAGE method was applied to simulate the dynamic response of structures under blast loads[56], demonstrating superior efficiency compared to traditional solvers. However, it faced challenges in long-term prediction accuracy and handling highly nonlinear phenomena. Finally, a GNN-LSTM model was employed to forecast vibration responses in frame structures induced by earthquakes[57]. This fusion framework effectively integrated structural graph embeddings with ground-motion sequences, achieving accurate nonlinear response predictions but being limited to specific geometries and numerically simulated training data.

Previous research primarily focused on developing dynamic digital twin models for individual structures, without considering those with diverse structural configurations. Furthermore, few studies tested the algorithms on real-world or experimental structures. Therefore, Graph-based Digital Twin Modelling is proposed to overcome the limited physical interpretability of other deep-learning methods and their lack of adaptability to diverse topological configurations and system parameters.

# 3. Methodology

## *3.1 The GDTM structural dynamics simulation method*

The fundamental procedure of the GDTM method for simulating structural dynamic response was illustrated in Figure 1. In this context, the intersection points of structural components were referred to as vertices $\boldsymbol{v}$, and the components connecting two vertices were referred to as edges $\boldsymbol{e}$. Consequently, any structure could be represented as a graph $\boldsymbol{G} = (\boldsymbol{v}, \boldsymbol{e})$. By assigning indices to the vertices and edges, the positional relationships between vertices and edges in a graph could be represented using an adjacency matrix $\mathbf{A}$, a process known as graph embedding or graph representation. The parameter $e_{ij}$ in the adjacency matrix $\mathbf{A}$ represented the weight of the edge between vertex $v_i$ and $v_j$ ; if an edge exists between the vertices, then $e_{ij} \neq 0$, otherwise $e_{ij} = 0$. $\boldsymbol{f}$ represented the feature vector of a vertex, typically containing the vertices' velocity and displacement responses, and the feature vectors of all vertices form the vertices feature matrix $\mathbf{F}$. By multiplying the adjacency matrix $\mathbf{A}$ with the vertices feature matrix, as shown in Figure 1(b), the feature vectors $\boldsymbol{f}$ of spatially adjacent vertices are extracted, multiplied by $e_{ij}$ , and summed, yielding $f_{in\text{-}} = \sum e_{ij} \cdot f_{jn}$, where the subscript $n$ represented the $n$-th feature of the vertex. This process was referred to as aggregation. The aggregated vertex features $f_{in\text{-}}$ were then input into a multilayer perceptron (MLP) along with the excitation signals $\boldsymbol{E}$, where the features of each vertex were updated, and the updated vertex features could subsequently be used as input to update the vertex features to the next state. The process of updating vertex features iterated until the stopping criterion was reached, as demonstrated in Figure 1(c).

$$\begin{aligned} \mathbf{u}_{n+1} &= \mathbf{u}_n + \Delta t \cdot \mathbf{v}_n + \frac{\Delta t^2}{2}\left((1-2\beta)\boldsymbol{a}_n + 2\beta\boldsymbol{a}_{n+1}\right) \\ \mathbf{v}_{n+1} &= \mathbf{v}_n + \Delta t\left((1-\gamma)\boldsymbol{a}_n + \gamma\boldsymbol{a}_{n+1}\right) \end{aligned} \tag{1}$$

To train the GDTM model, acceleration responses from numerical simulations or measurements, along with the external excitation time history signals applied to each vertex, were sampled at a fixed frequency. The acceleration data were then numerically integrated to derive the corresponding displacement and velocity responses, as shown in Equation (1). For Equation (1), $\mathbf{u}$, $\mathbf{v}$, and $\mathbf{a}$ represented the displacement response, velocity response, and acceleration response, respectively. The subscript $n$ indicated the time step. $\Delta t$ denoted the sampling time interval, which is the reciprocal of the sampling frequency. $\beta$ and $\gamma$ were parameters of the numerical integration method, typically chosen as $\gamma \geq 0.5$, $\beta \geq 0.25(0.5+\gamma)^2$. The collected data were subsequently divided into input and output sets: the input data consisted of the velocity $\boldsymbol{v}$ and displacement $\boldsymbol{u}$ responses at time step $n$ for each vertex, as well as the excitation signal $\boldsymbol{E}$ at time step $n$+1; the output data comprised the acceleration responses $\boldsymbol{a}$ of each vertex at time $n$+1.

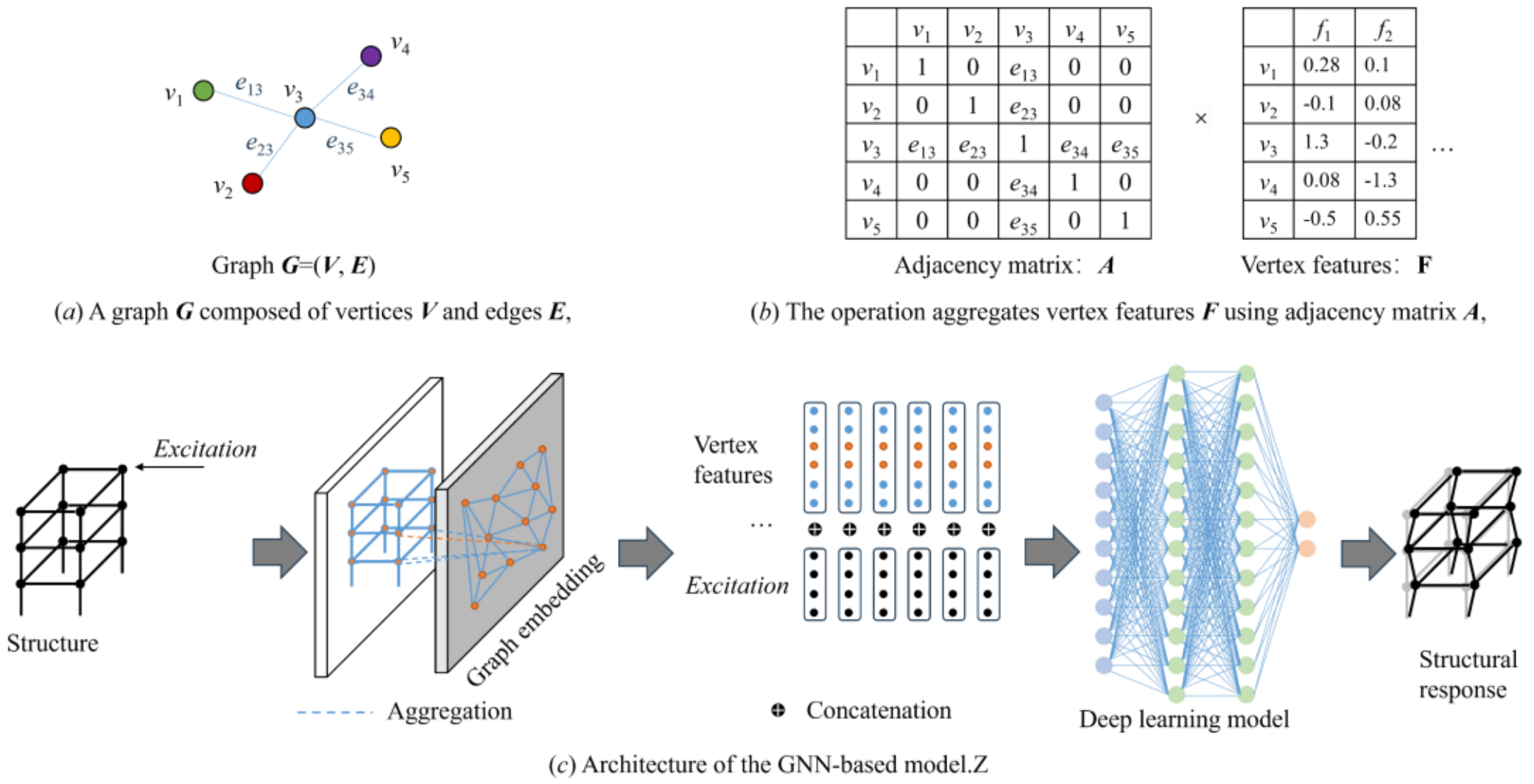

**Figure 1.** The basic process diagram of the GDTM structural dynamic simulation method.

To address different structural forms, this study also proposed homogeneous and heterogeneous GNN-based methods. The choice between homogeneous GNN-based and heterogeneous GNN-based methods primarily depended on whether the components making up the structure were of the same type. If all the components in the structure, had the same properties, a homogeneous graph was used to represent the structure. The adjacency matrix of the homogeneous graph was as illustrated in Figure 2(a). Conversely, multiple adjacency matrices were required for different components, as shown in Figure 2(b). The number of adjacency matrices was equal to the number of component types plus one, with the additional adjacency matrix used to aggregate the attributes of each vertex itself.

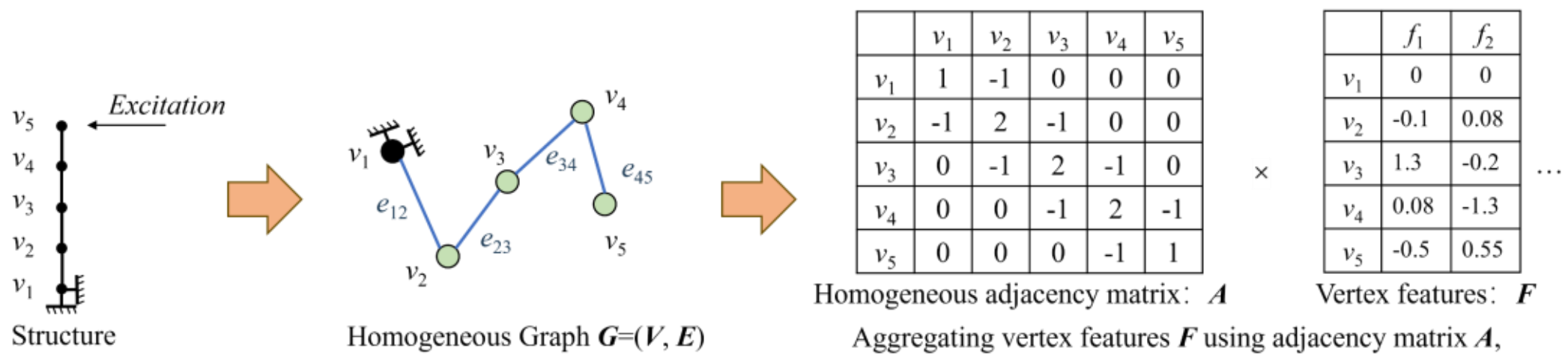

($a$) Homogeneous graph and its adjacency matrix.

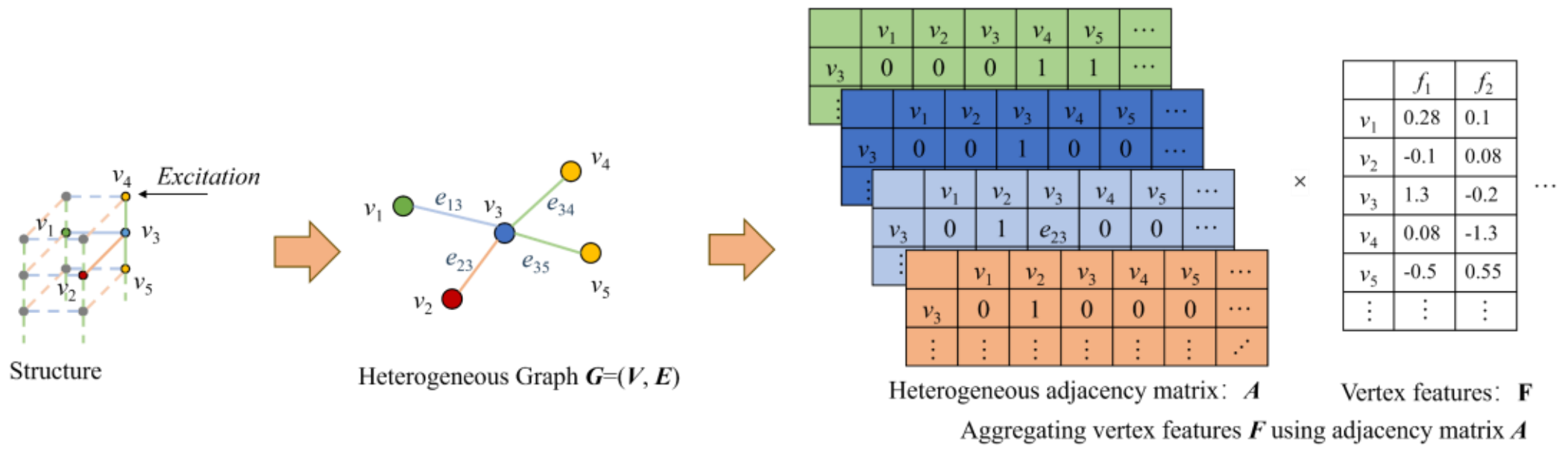

($b$) Heterogeneous graph and its adjacency matrix

**Figure 2.** Comparison of homogeneous and heterogeneous graphs.

One important challenge in this study was determining how to construct an adjacency matrix. Since different components exhibited varying characteristics, the elements in the adjacency matrix varied accordingly. To address this, an attention mechanism was employed to dynamically update these element values, offering a potential solution. In this investigation, the attention mechanism employed in GAT[50]. Therefore, this approach was referred to as the GAT-based digital twin modeling method (GAT-based method).

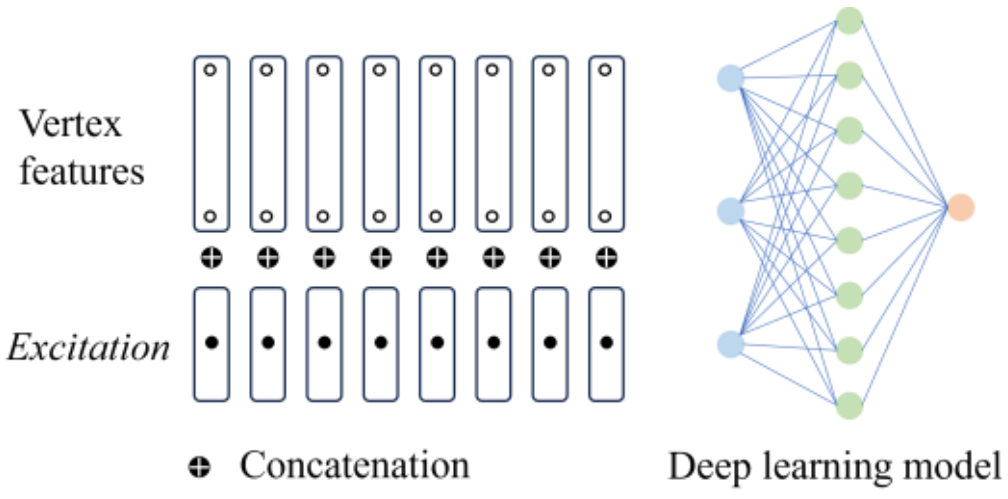

($a$) Homogeneous graph.

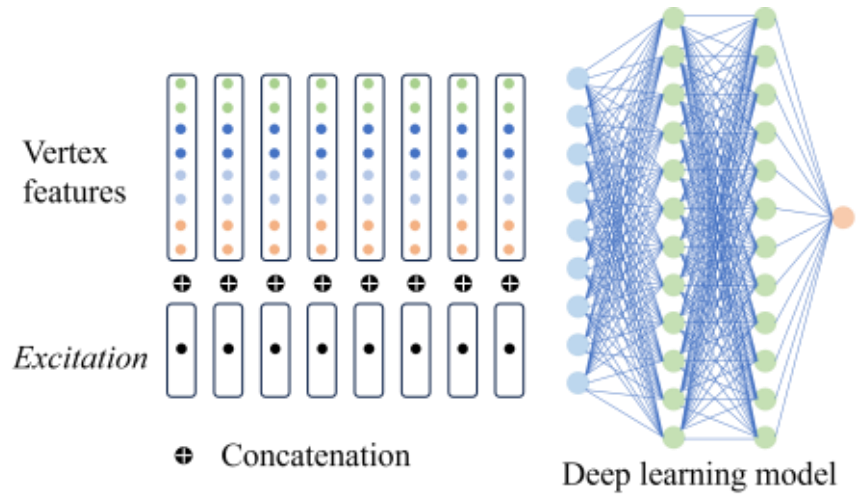

($b$) Heterogeneous graph.

**Figure 3.** Vertex feature update in GDTM Methods.

Figure 3 illustrates the process of vertex feature updates in the GDTM structural dynamics simulation methods. The homogeneous GNN-based and GAT-based methods employed a single adjacency matrix to aggregate the features of all vertices. The aggregated vertex features were then concatenated with the excitation signals to form the input data, which had a dimensionality of 3, as shown in Figure 3(a). On the other hand, heterogeneous GNN-based method utilized $N$ distinct adjacency matrices, resulting in a final input data dimensionality of $2N$+1, as depicted in Figure 3(b).

## *3.2 Model training and GDTM structural dynamic response simulation*

In essence, the training process of the GDTM method followed a similar procedure for both homogeneous GNN-based, heterogeneous GNN-based, and GAT-based methods. These collected acceleration data, along with the numerically integrated velocity and displacement response data and the recorded excitation signals, were then split into input data **Input**=($\hat{\boldsymbol{v}}_n$, $\hat{\boldsymbol{u}}_n$, $\hat{\boldsymbol{E}}_{n+1}$) and output data **Output** =($\hat{\boldsymbol{a}}_{n+1}$), where n=1,2,…,t represents the sampled time steps. Following a series of steps such as graph embedding, adjacency matrix aggregation, and vertex feature updates, the model produced the predicted vertex acceleration response for the subsequent time step. The complete procedure was detailed in Algorithm 1.

**Algorithm 1** Training GNN model

**Require:** **G**=($\boldsymbol{v}$, $\boldsymbol{e}$). $\mathbf{X}_i$: vertex features(e.g., velocity, displacement) at time $i$ , $i$=0,1,…,T-1. $\boldsymbol{E}_i$ : Excitation time history data. **A**: adjacency matrix. **GNN-MODEL**: Initial GNN model

**Ensure:** Calculate the acceleration response $\boldsymbol{a}_{i+1}$ of vertexes

1: **Input data**←($\mathbf{X}_i$, $\boldsymbol{E}_{i+1}$) , $i$=0,1,…,T-1

2: **for** vertex $v$ ∈ $\boldsymbol{v}$ **do**

3: $\boldsymbol{f}_{\text{in-}}$←Aggregate({**A** ×$\mathbf{X}_i$})

4: $\boldsymbol{h}_{\text{in}}$←Concatenate($\boldsymbol{f}_{\text{in-}}$, $\boldsymbol{E}_{i+1}$)

5: $\boldsymbol{a}_{i+1}$←**GNN-MODEL**($\boldsymbol{h}_{\text{in}}$)

6: **end for**

7: **return** $\boldsymbol{a}_{i+1}$, $i$=0,1,…,T

Algorithm 2 illustrates the GDTM structural dynamics simulation method proposed in this study. Only the initial conditions, often set to zero, and the excitation time history data were available during the dynamic response simulation. Afterwards, the algorithm numerically integrated the acceleration response to compute the vertex features, specifically the velocity response and displacement response, as shown in Equation (1). These features were then fed into the MLP along with the excitation data in the subsequent iteration to compute the acceleration response of each vertex at the next time step. This process was repeated until the algorithm's stopping criterion was met.

**Algorithm 2** GNN-based Structural Dynamics Simulation

**Require:** **G**=($\boldsymbol{v}$, $\boldsymbol{e}$). $\mathbf{X}_0$: initial vertex features (e.g., velocity, displacement). $\boldsymbol{E}_i$ : Excitation time history data, i=0,1,…,T-1. **A**: adjacency matrix. **GNN-MODEL**: trained GNN model

**Ensure:** Calculate the acceleration response $\boldsymbol{a}_i$ of vertexes at the each time step.

1: Initialize **Input data**←($\mathbf{X}_0$, $\boldsymbol{E}_1$)

2: **for** t=1 to T **do**

3: **for** vertex $v$ ∈ $\boldsymbol{v}$ **do**

4: $\boldsymbol{f}_{\text{in-}}$←Aggregate({**A** ×$\mathbf{X}_t$})

5: $\boldsymbol{h}_{\text{in}}$←Concatenate($\boldsymbol{f}_{\text{in-}}$, $\boldsymbol{E}_{t+1}$)

6: $\boldsymbol{a}_{t+1}$←**GNN-MODEL**($\boldsymbol{h}_{\text{in}}$)

7: $\mathbf{X}_{t+1}$←Numerical Integration($\boldsymbol{a}_{t+1}$)

8: **end for**

9: **end for**

10: **return** $\boldsymbol{a}_i$, $i$=0,1,…,T-1

During the training process, a variant of the Adam algorithm for stochastic gradient descent was employed as the optimization method. The Adam stochastic gradient descent algorithm, combined with the Smooth-L1 loss function, was used as the optimizer and loss function, respectively. The key parameters for the Adam optimizer were set as follows: the learning rate was $3 \times 10^{-4}$ , $\beta_1$ was 0.9, $\beta_2$ was 0.999.

# 4. Results and discussion

To validate the proposed method, both numerical simulations and laboratory experiments were conducted to assess its performance and applicability. The numerical model employed was M-DOF systems. In the laboratory experiments, a modular steel building with a planar frame structure was used to evaluate the method's effectiveness.

## *4.1 Numerical case*

This section focuses on simulating the dynamic responses of the M-DOF systems, primarily discussing the training process of the homogeneous GNN and its accuracy. Furthermore, the simulation of dynamic responses for M-DOF systems with diverse topological structures and different system parameters using this method is also discussed.

### *4.1.1 Data preparation and model training*

First, a 10-DOF system was established, as shown in Figure 4. Each node had only a translational degree of freedom in the *x* direction, with adjacent nodes connected by springs and dampers. The mass of each node, as well as the stiffness and damping between adjacent nodes, were set to uniform values: the mass of each node was 2000 kg, the stiffness between nodes was $2.4 \times 10^5$ N/m, and the damping was 2500 $N{\cdot}s/m$. A free-body diagram force analysis was performed on each node using Newton's second law, as Equation (2), thereby establishing the 10-DOF numerical model. The $Newmark{-}\beta$ numerical integration method was then employed to calculate the acceleration response$\ddot{\mathbf{Y}}(t)$, velocity response $\dot{\mathbf{Y}}(t)$, and displacement response $\mathbf{Y}(t)$ under three types of commonly used single-node excitations: impulse excitation, harmonic excitation, and random excitation.

In Equation (2), $\mathbf{F}(t)$ denoted the time-varying excitation, while $\mathbf{L}$ was the excitation influence matrix responsible for applying the excitation influence matrix responsible for applying the excitation to the structure. The matrices $\mathbf{M}$, $\mathbf{C}$, and $\mathbf{K}$ corresponded to the system's mass, damping, and stiffness, respectively. The parameters *M*, *C*, and *K* represent the parameters of the system.

$$\mathbf{M} \cdot \ddot{\mathbf{Y}}(t) + \mathbf{C} \cdot \dot{\mathbf{Y}}(t) + \mathbf{K} \cdot \mathbf{Y}(t) = \mathbf{L} \cdot \mathbf{F}(t) \tag{2}$$

The node numbers in this model were assigned from 1 to 10, from bottom to top. The structural response acceleration of all nodes was collected at a sampling frequency of 100 Hz over a 50-second duration. For each type of excitation, the model's response was recorded 10 times. The recorded excitation data and the model's response were then split into training and test sets with an 8:2 ratio. Before training the model, it is necessary to standardize the input and output data. In this study, the data were normalized by dividing by the maximum absolute value of the response data. The MLP network architecture used in the GDTM model, along with its parameters, is shown in Table 1. The loss history curve during the training of the surrogate model is presented in Figure 5 .

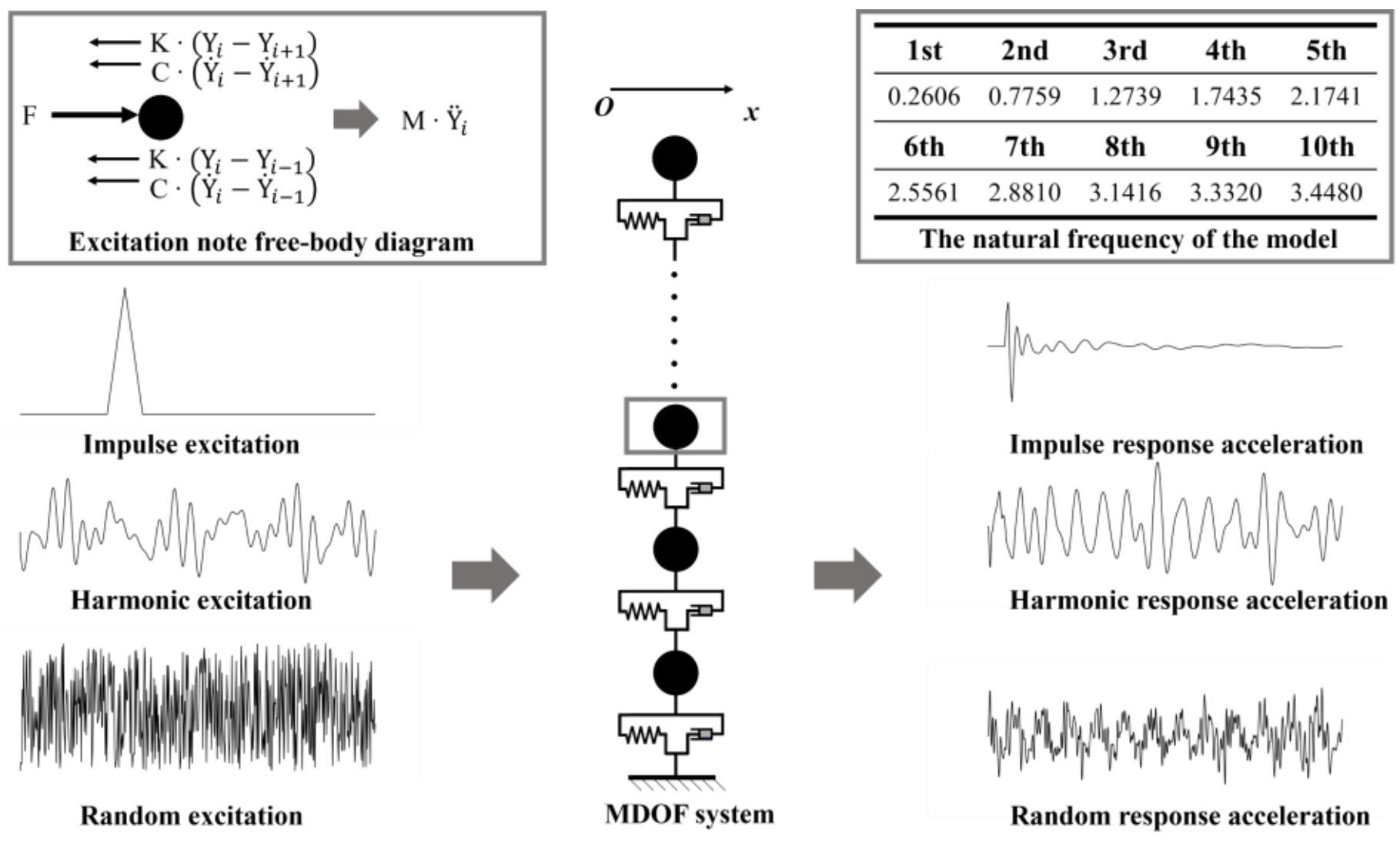

**Figure 4.** Schematic Diagram of the Numerical M-DOF System.

**TABLE 1** The configuration of the MLP architecture.

| Layer | Type | Activation | Input shape | Output shape |
|---|---|---|---|---|
| 1 | Linear | ReLU | 3 | 16 |
| 2 | Linear | ReLU | 16 | 64 |
| 3 | Linear | — | 64 | 1 |

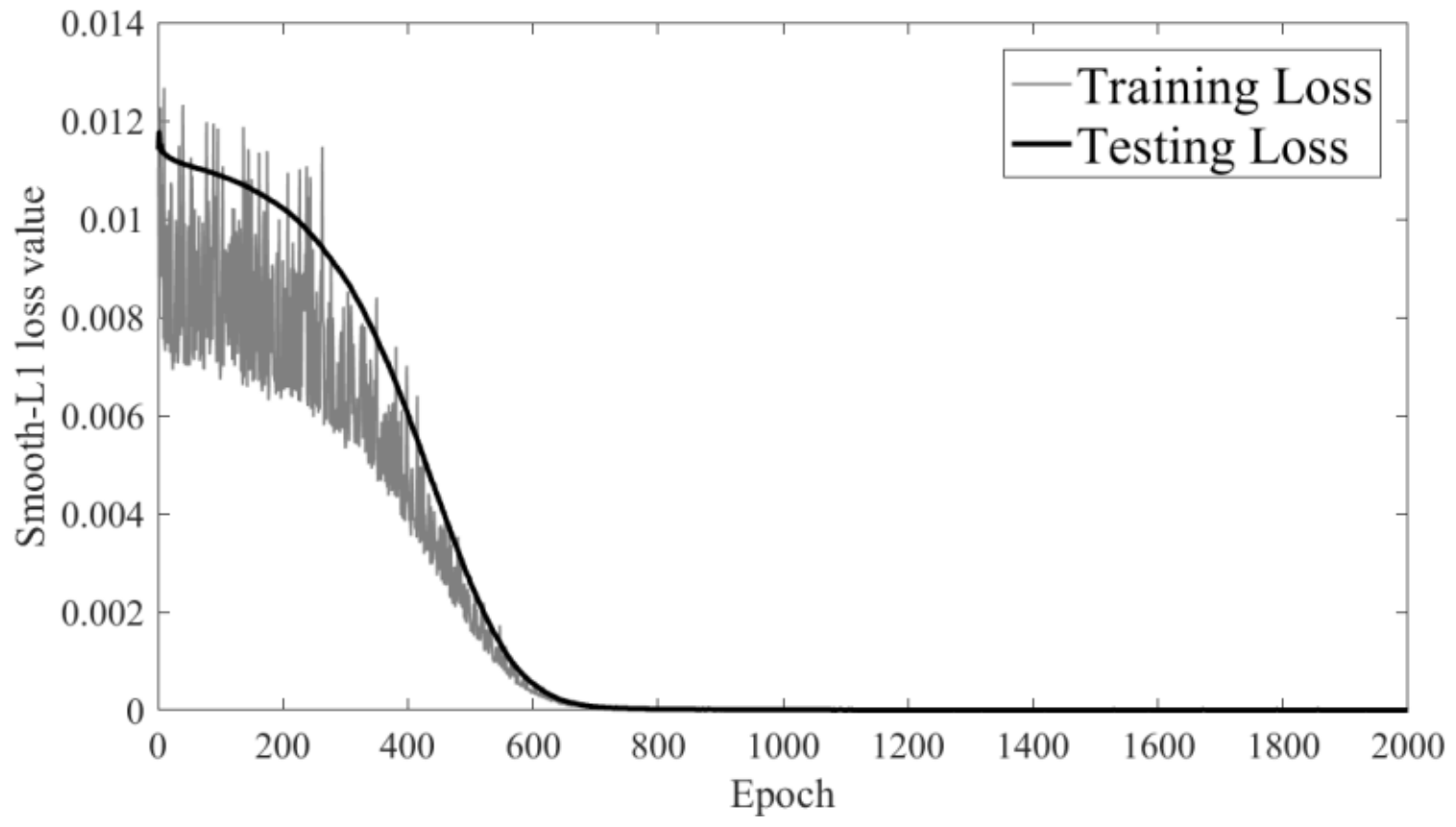

**Figure 5.** Loss history of the train dataset and the test dataset during training the surrogate modeling phase.

*4.1.2 Simulation results*

To evaluate the effectiveness of the GDTM structural dynamics simulation method, three evaluation metrics were established, as shown in Equation (3) to (5). The Normalized Mean-Squared Error (NMSE), the Coefficient of Determination ($R^2$), and Peak Error (%) of the GDTM method simulation results were $5.02\times10^{-3}$, 0.99, and 0.86%, respectively. According to the numerical results, the method introduced in this study demonstrated minimal error in the simulation of the system's acceleration response. The overall fluctuation pattern of the simulated acceleration closely aligned with the actual values, and the simulated outliers remained below 1%.

$$\mathbf{NMSE} = \frac{\sum_i^n \left(y_{true}^i - y_{pred}^i\right)^2}{\sum_i^n \left(y_{true}^i\right)^2} \tag{3}$$

$$\mathbf{R^2} = \mathbf{1} - \frac{\sum_{i=1}^n \left(y_{true}^i - y_{pred}^i\right)^2}{\sum_{i=1}^n \left(y_{true}^i - \boldsymbol{mean}\left(y_{pred}^i\right)\right)^2} \tag{4}$$

$$\mathbf{PE}(\%) = \frac{\boldsymbol{max}\left(\left|y_{true}^i - y_{pred}^i\right|\right)}{\boldsymbol{max}\left(\left|y_{true}^i\right|\right)} \times 100 \tag{5}$$

Partial results of the acceleration response simulated by the GDTM method were shown in Figures 6. The figure illustrates the simulated acceleration responses and ground truth of the system at DOF 1, 3, 5, 6, 8, and 10 under three different load conditions. The results demonstrate that the GDTM structural dynamics simulation method can accurately simulate the structural response of the 10-DOF system under various dynamic loads, validating the effectiveness of this approach.

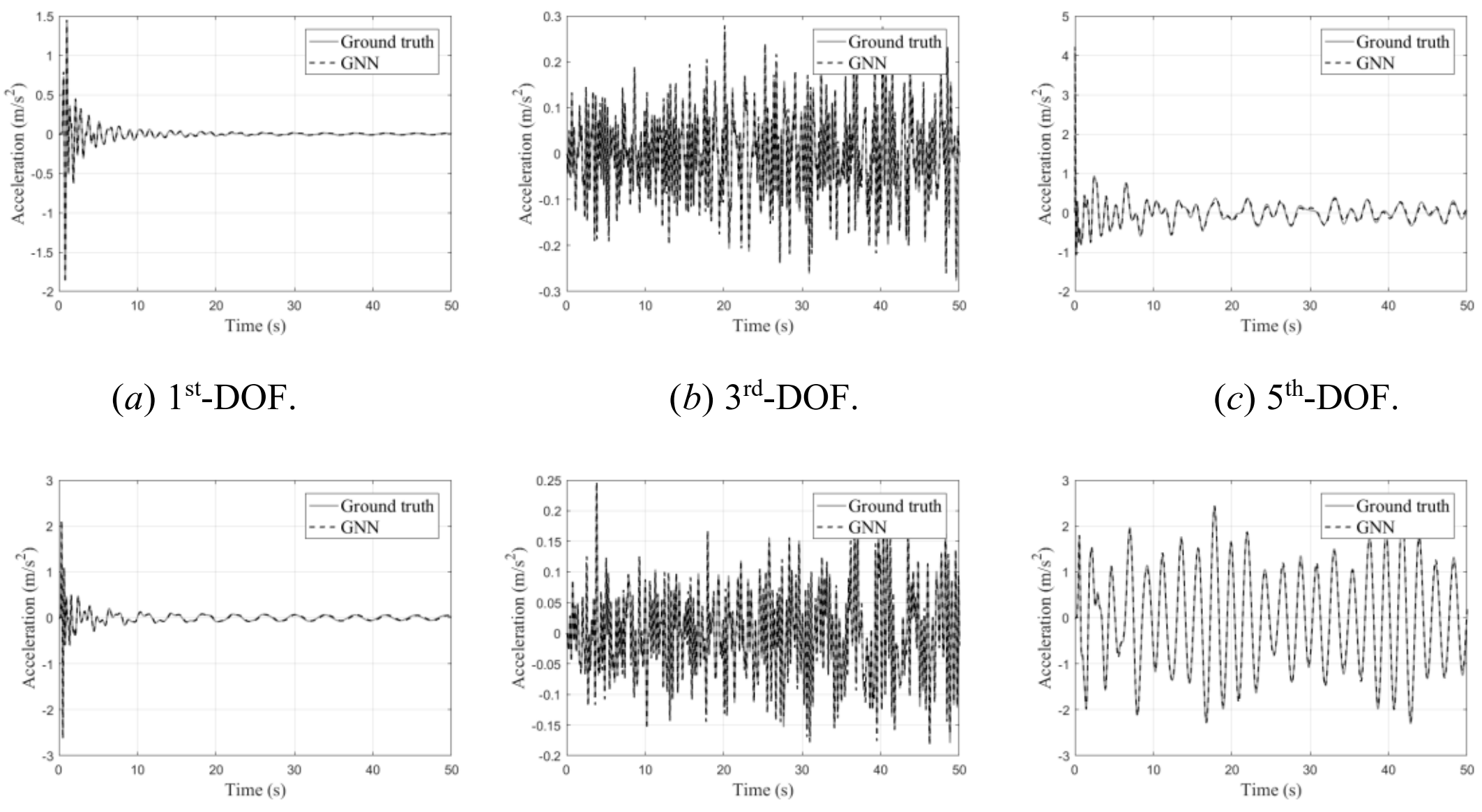

($a$) 1st-DOF. ($b$) 3rd-DOF. ($c$) 5th-DOF.

($d$) 6th-DOF. ($e$) 8th-DOF. ($f$) 10th-DOF.

**Figure 6.** Partial Results of the GNN Model Simulated 10-DOF system Acceleration Response.

*4.1.3 Verifying GDTM model's ability to simulate structural response with different topologies*

GNN belong to inductive learning, characterized by their ability to infer global features from local ones. To further verify the GDTM model's capability to simulate different spatial topological structures, four distinct M-DOF systems were established: 5-DOF, 12-DOF, 20-DOF, and 30-DOF. The mass, damping, and stiffness parameters of these new M-DOF systems were identical to those of the 10-DOF system. The GDTM model, trained on the 10-DOF system data, successfully simulated the responses of the four new M-DOF systems by merely adjusting the adjacency matrix, without modifying any model parameters.

**TABLE 2** Three evaluation metrics value of GNN model simulation the response of system with different topologies.

| | 5-DOF | 12-DOF | 20-DOF | 30-DOF |
|---|---|---|---|---|
| NMSE | 0.013 | $1.78\times10^{-4}$ | $2.55\times10^{-4}$ | $2.95\times10^{-4}$ |
| $R^2$ | 0.90 | 0.99 | 0.99 | -0.093 |
| PE (%) | 5.80 | 1.92 | -0.21 | -0.70 |

The simulation results of the GDTM model for the dynamic responses are presented in Table 2. From the results in Table 2, it was evident that, except for the 5-DOF system, the GDTM model trained on 10-DOF system data successfully simulated systems with different topological structures, provided the corresponding adjacency matrix was clearly defined. To visualize the GDTM model's performance, the simulated acceleration response of the 5-DOF system was displayed in Figure 7 . As shown in Figure 7 , although the GDTM model exhibited a larger error in simulating the 5-DOF system, this discrepancy arose because the 5-DOF system's response was an order of magnitude smaller compared to other systems. Nonetheless, the simulated results were still largely consistent with the ground truth, confirming that this method effectively captured the dynamic responses of systems with various topological forms.

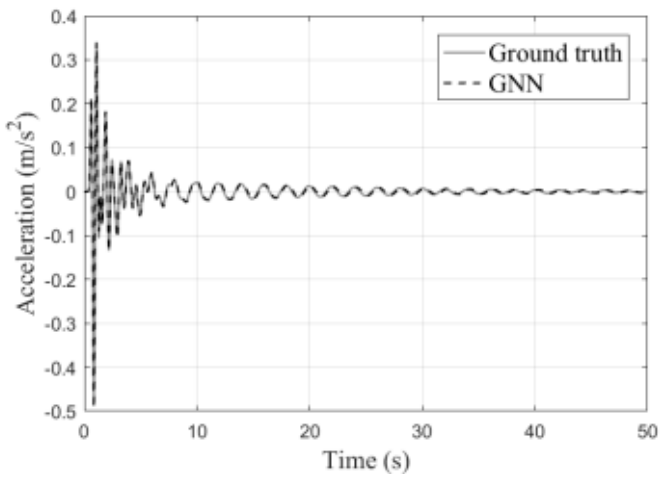

(*a*) 1st-DOF.

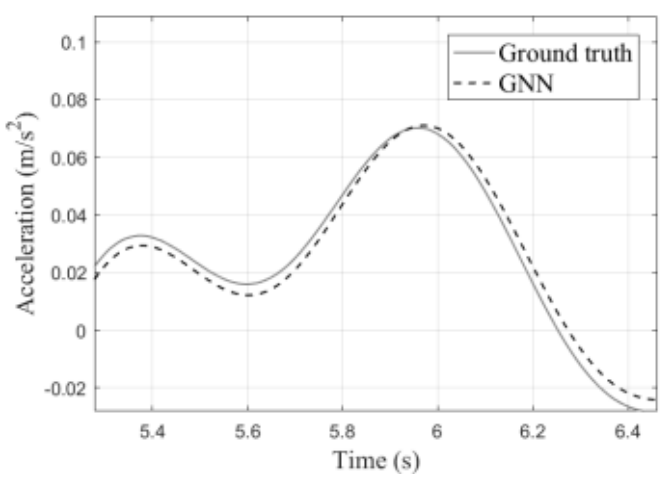

(*b*) 2nd-DOF.

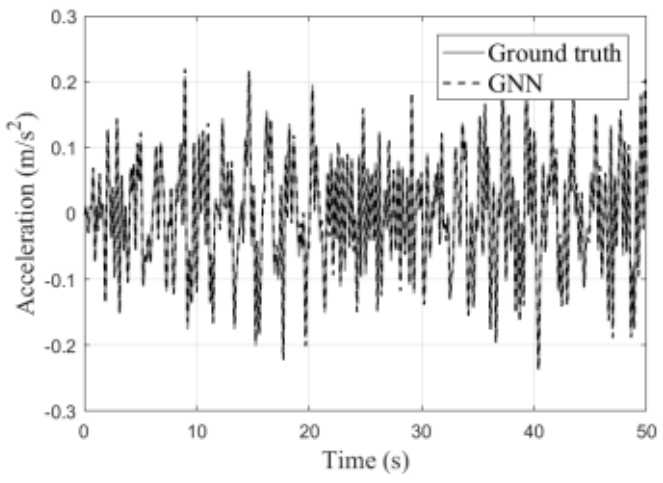

(*c*) 3rd-DOF.

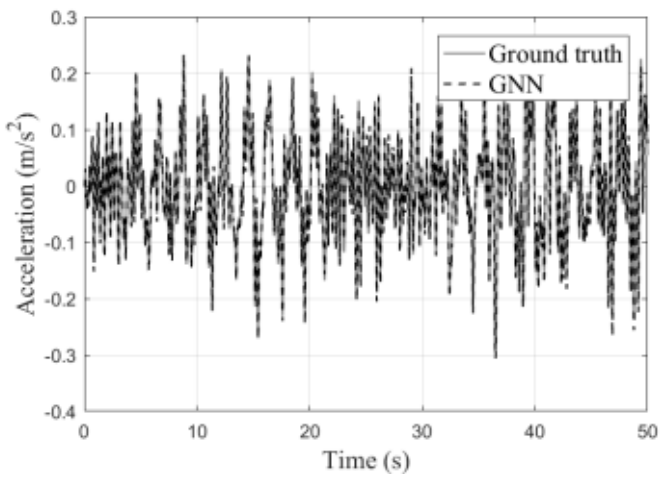

(*d*) 4th-DOF.

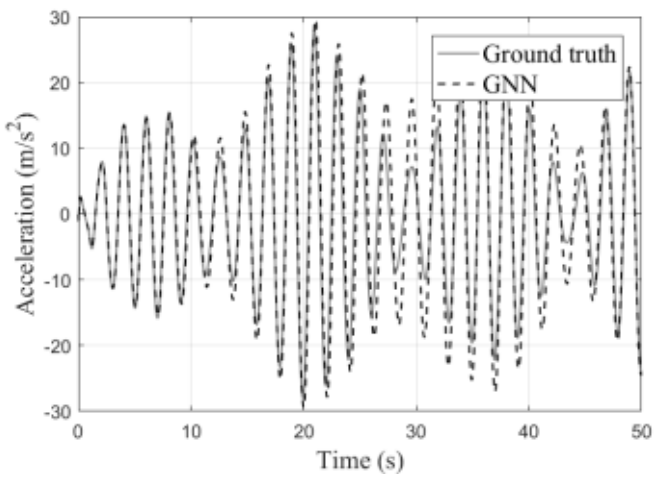

(*e*) 5th-DOF.

**Figure 7 .** Results of the GNN Model Simulated 5-DOF system Acceleration Response.

*4.1.4 Verifying GDTM model's ability to simulate structural response with different system parameters*

To further clarify the role of the adjacency matrix in GDTM models and assess the method's ability to simulate the dynamic response of various systems, this subsection explored the modification of the adjacency matrix values to simulate the dynamic response of 10-DOF systems with different system parameters. To accomplish this, in addition to the original 10-DOF system (CASE 0), three distinct 10-DOF systems were constructed by altering the system parameters, labeled as CASE 1, CASE 2, and CASE 3. The parameters for all four systems are detailed in Table 3.

**TABLE 3** 10-DOF system configurations.

| | **CASE 0** | **CASE 1** | **CASE 2** | **CASE 3** |
|---|---|---|---|---|
| Stiffness | K | 0.8×K | 0.1×K | Rand×K |
| Mass | M | 1.6×M | M | Rand×M |
| Damping | C | C | 0.1×C | Rand×C |

In Table 3, for CASE 1 and CASE 2, the mass (M), damping (C), and stiffness (K) matrices were uniformly scaled either up or down for the entire system. CASE 3, however, employed random values ranging from 0.5 to 1.5 to introduce variability in the parameters across different DOF.

**TABLE 4** Three evaluation metrics value of simulated response of 10-DOF system with different parameters.

| | CASE 0 | CASE 1 | CASE 2 | CASE 3 |
|---|---|---|---|---|
| NMSE | $5.02\times10^{-3}$ | 0.38 | 0.015 | 0.082 |
| $R^2$ | 0.99 | 0.62 | 0.99 | 0.92 |
| PE (%) | 0.86% | 57.01% | 22.86% | 27.09% |

The corresponding adjustments in the adjacency matrix were made through proportional scaling at the relevant positions. The resulting simulation outcomes are displayed in Table 4. Of all the cases, CASE 1, in particular, displayed the largest deviation. To offer a clearer visual representation of the gap between these results and the actual values, selected results from CASE 1 were presented in Figure 8 . Based on the results shown in Figure 8 , the fluctuation patterns of the GDTM model's simulation closely aligned with the ground truth.

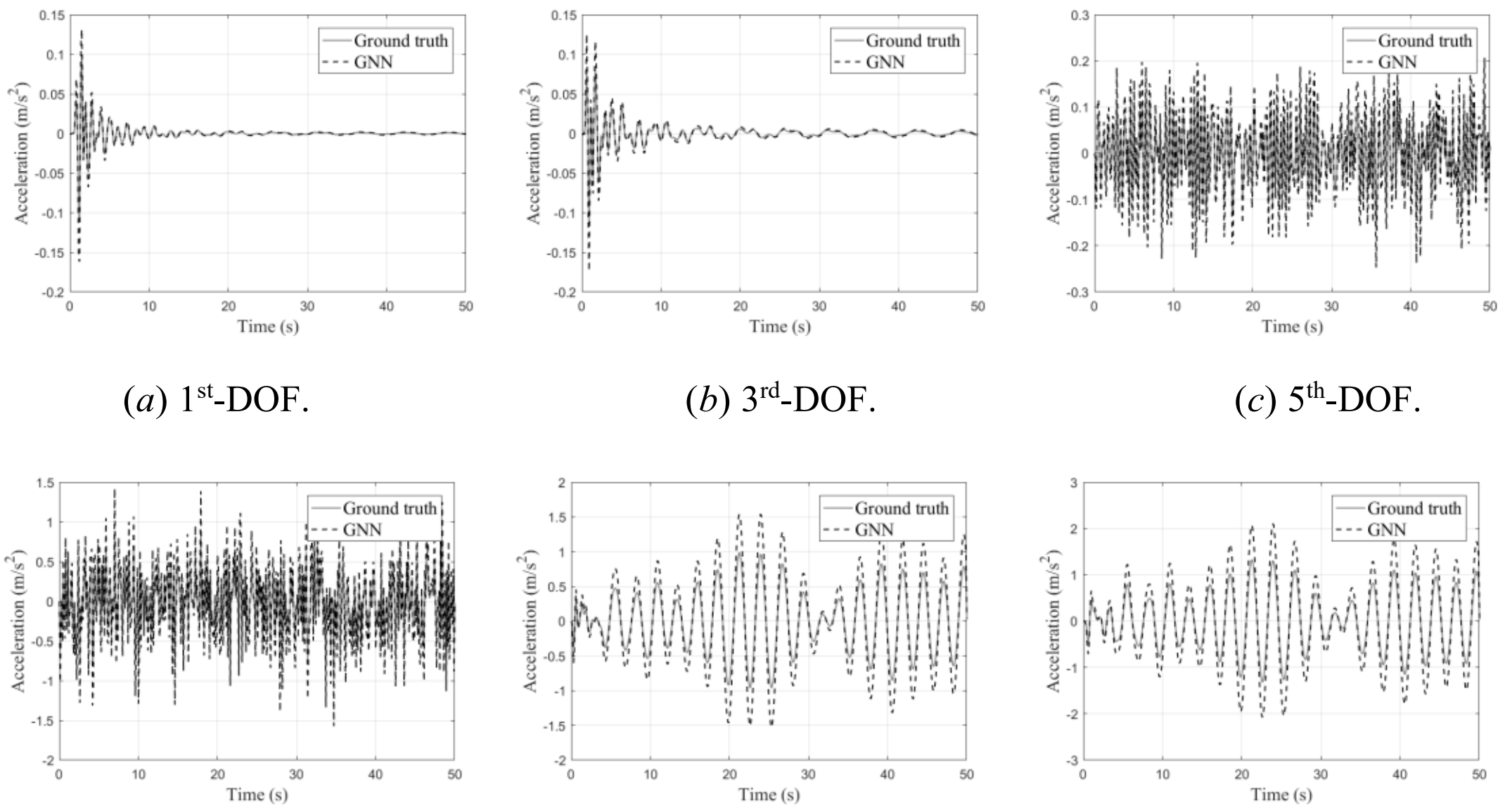

(*a*) 1st-DOF. (*b*) 3rd-DOF. (*c*) 5th-DOF.

(*d*) 6th-DOF. (*e*) 8th-DOF. (*f*) 10th-DOF.

**Figure 8 .** Partial Results of the GNN Model Simulated CASE 1 Acceleration Response.

## *4.2 Experimental case*

To validate the GDTM method's capability to accurately simulate real-world structural systems, it was demonstrated that the GDTM structural dynamics simulation not only surpassed traditional approaches in simulation accuracy but also significantly improved computational efficiency. In the laboratory, three modular steel building frame models were constructed, representing 2-storey (2S), 3-storey (3S), and 4-storey (4S) structures, as shown in Figure 9 .

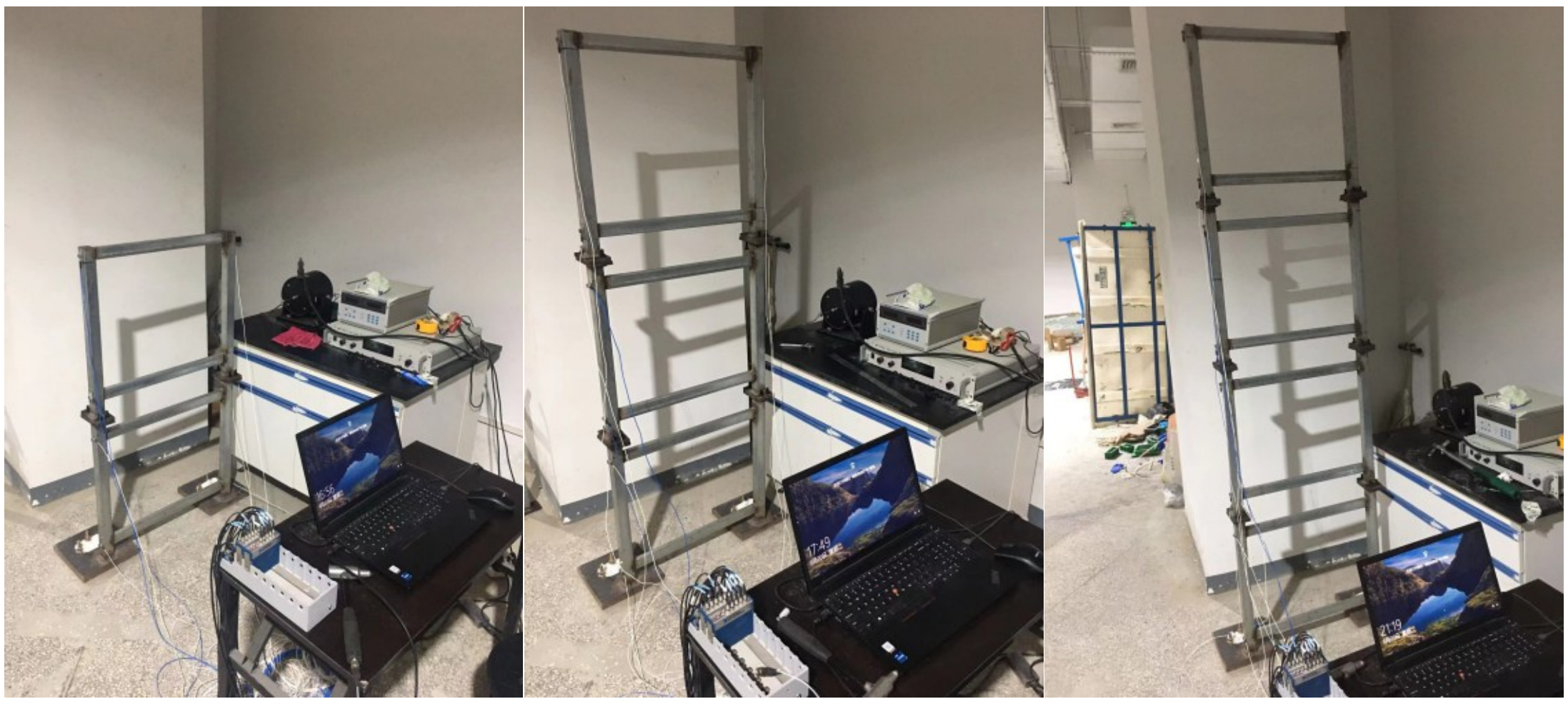

(*a*) 2-storey model(2S). (*b*) 3-storey model (3S). (*c*) 4-storey model(4S).

**Figure 9.** Modular steel buildings experimental models.

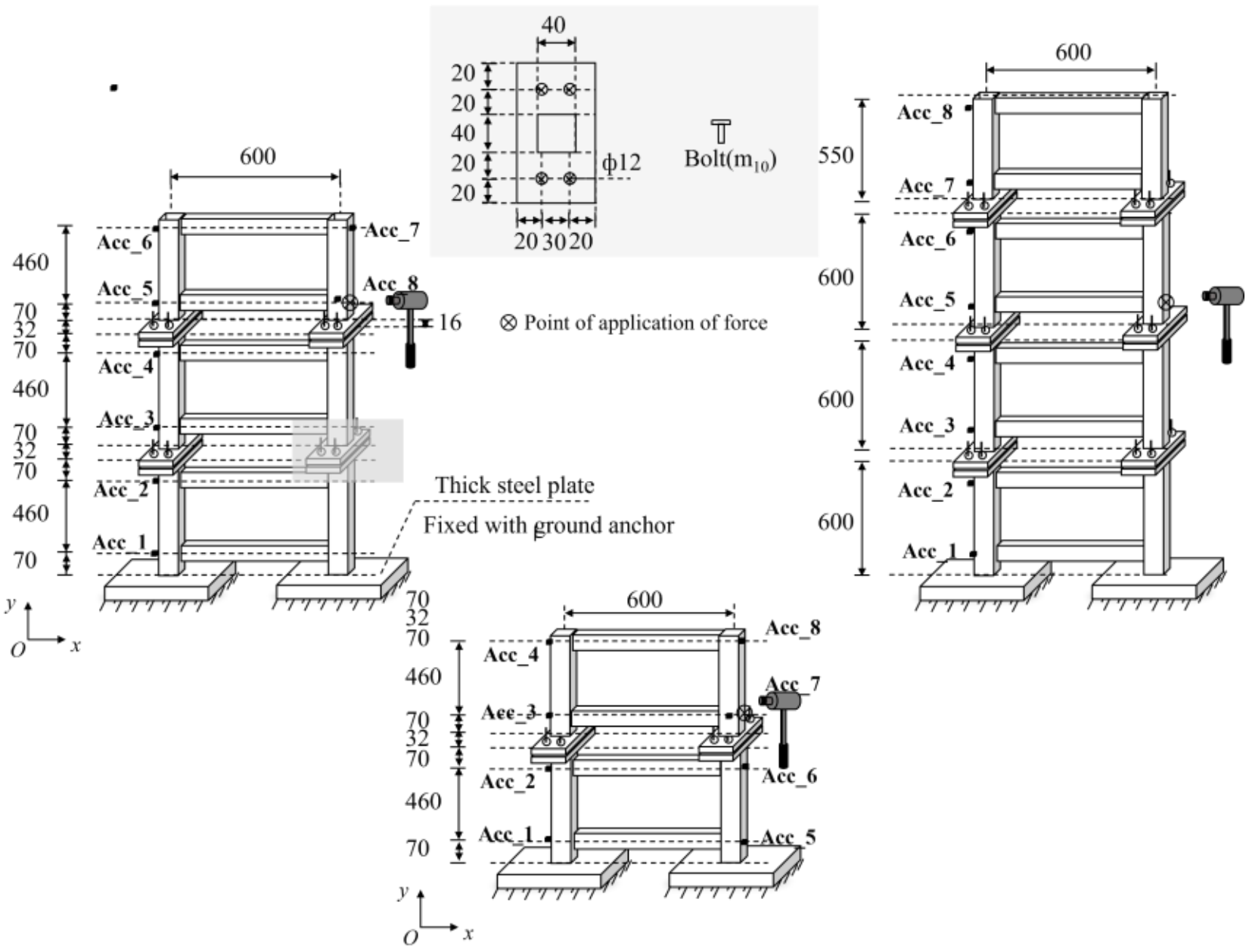

**Figure 10.** Experimental model diagram. Unit: mm.

**TABLE 5** Cross section dimensions and material parameters of beams and columns. Where, SHS: square hollow sections, RHS: rectangular hollow sections. Unit: mm.

| Key | Material quality | Column section | Beam section | Young's modulus (MPa) | Density (kg/m$^3$) | Poisson's ratio |
|---|---|---|---|---|---|---|
| Value | Q235 | 40 × 40 × 2 **SHS** | 20 × 40 × 2 **RHS** | $2.06 \times 10^5$ | 7800 | 0.25 |

The bases of these experimental models were welded to expansion bolts, which were securely embedded into the ground. The detailed dimensions of the experimental models, along with the positions of the acceleration sensors and hammer impact locations, were illustrated in Figure 10. The cross-sectional dimensions and material properties of the beams and columns were listed in Table 5.

*4.2.1 Data preparation and model training*

The vibration test of the modular steel frame experimental structures was conducted under impulse excitation generated by an impact hammer. The hammer, a PCB-086D05 model, delivered a peak force of 22240N. Six unidirectional accelerometers, the PCB-333B32 model, were used to measure the vibration response of the model. A data acquisition system, specifically the NI 9231 model, was used to simultaneously capture both the acceleration responses and the impulse excitations. The setup for the vibration test is shown in Figure 1 1 . In total, 20 separate vibration tests were performed using impulse excitation, following the guidelines of ISO 7626-5. The sampling rate was set to 5120 Hz to ensure sufficient modal information, with each test lasting 12 seconds. To enhance the data utilization, the original dataset was down-sampled by selecting one point every 10 samples, which increased the dataset size by a factor of ten, resulting in 200 datasets. The data were then split into training and testing subsets with an 8:2 ratio.

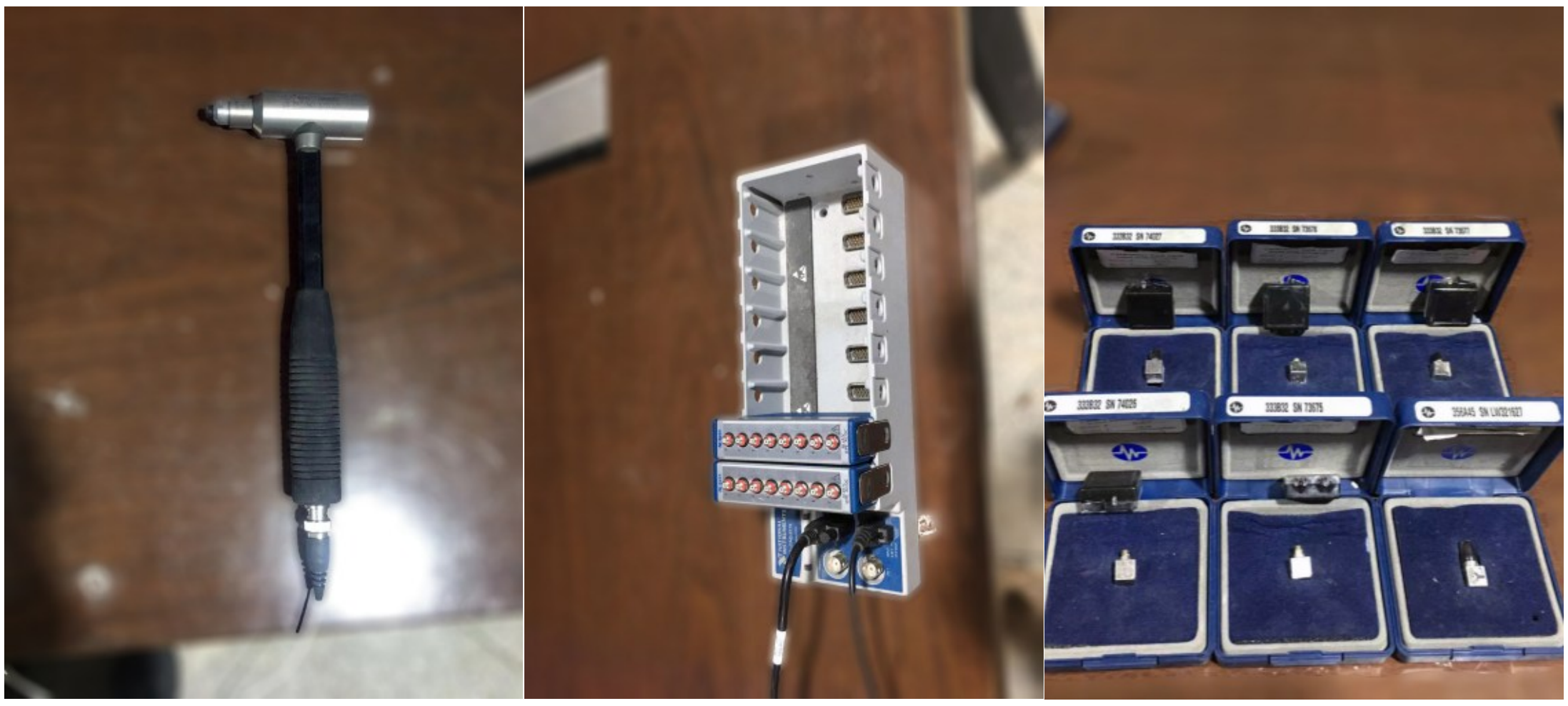

(*a*) Impact Hammer. (*b*) Acquisition card. (*c*) Acceleration sensors.

**Figure 1 1 .** The vibration testing equipment.

To accurately capturing the dynamic behavior of diverce components, heterogeneous GNN-based and GAT-based methods were introduced to simulate the dynamic response of structures consisting of various component types. The detailed procedures for training and simulating the structural dynamic response using the heterogeneous GNN-based and GAT-based methods were presented in Algorithm 1 and Algorithm 2, respectively. Further specifics were discussed in Section 3 of this paper.

*4.2.2 Simulating dynamic response of 2S structure*

Data collected from the acceleration sensors of the 2S structure were utilized to train the heterogeneous GNN-based and GAT-based models. The dynamic response of the two-story steel frame under impact loads was then simulated, with the results detailed in Table 6.

**TABLE 6** Three evaluation metrics value of GNN-based model and FEM.

| GNN | | | GAT | | | FEM | | |
|---|---|---|---|---|---|---|---|---|
| NMSE | $R^2$ | PE (%) | NMSE | $R^2$ | PE (%) | NMSE | $R^2$ | PE (%) |
| 0.0015 | 0.9984 | 7.93 | 0.0038 | 0.9960 | 15.64 | 3.95 | 0.5212 | 192.97 |

To compare with the GDTM methods, a numerical simulation model of the experimental structure was developed using FEM, with the detailed construction process outlined in the Appendix. As indicated in Table 6, the comparison results revealed that the GDTM model surpassed the FEM numerical model in all acceleration response metrics, with the heterogeneous GNN-based model delivering the most optimal performance. To offer a more intuitive comparison of the differences between the simulated and measured acceleration responses, partial acceleration responses for parts of the structure, simulated using three different methods, are provided in Figures 12 and 13.

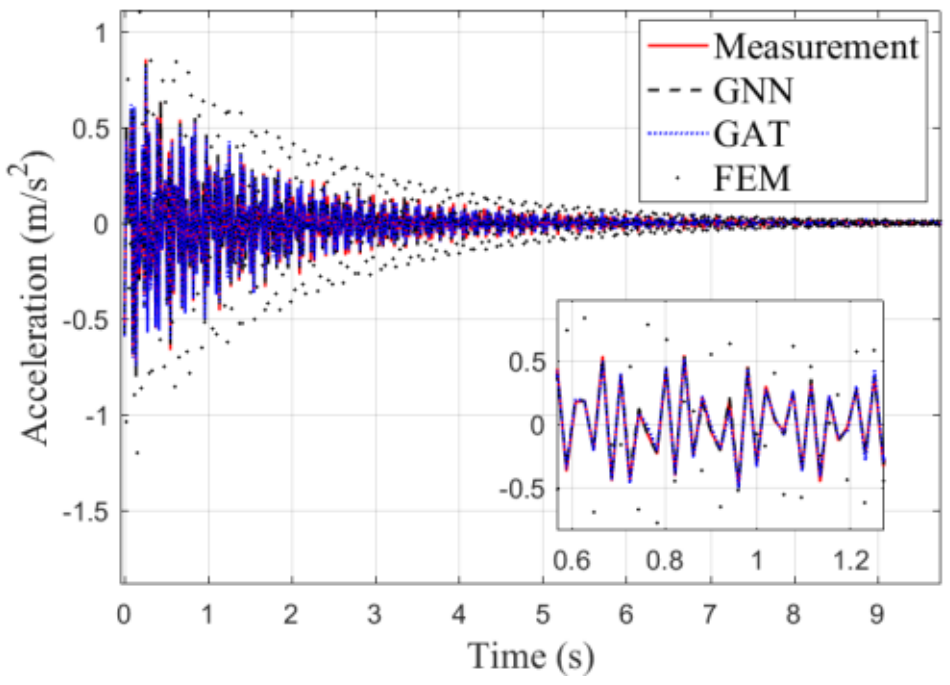

(*a*) Acc_1.

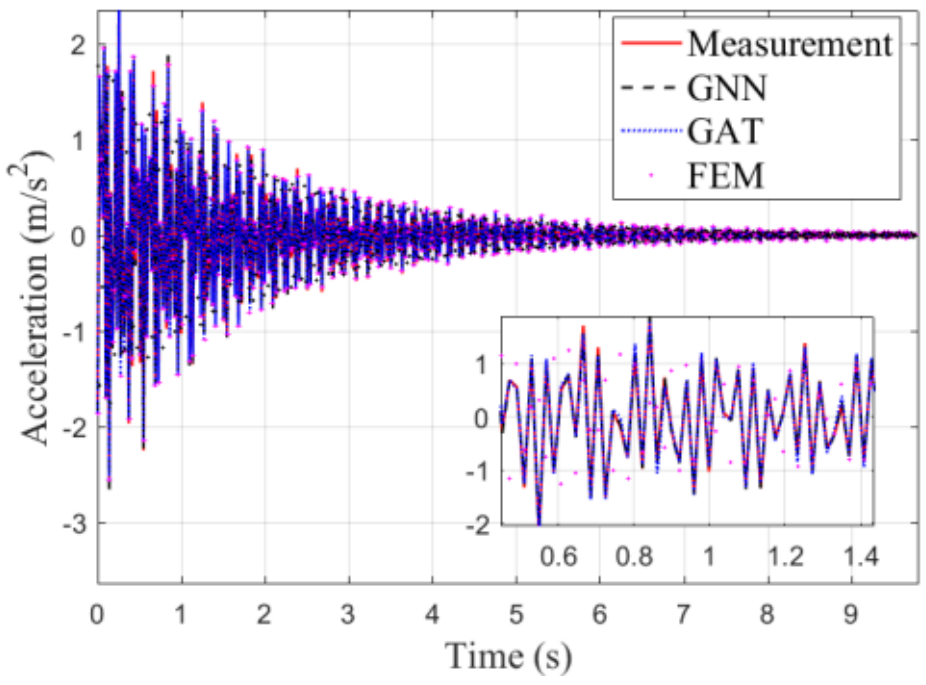

(*b*) Acc_2.

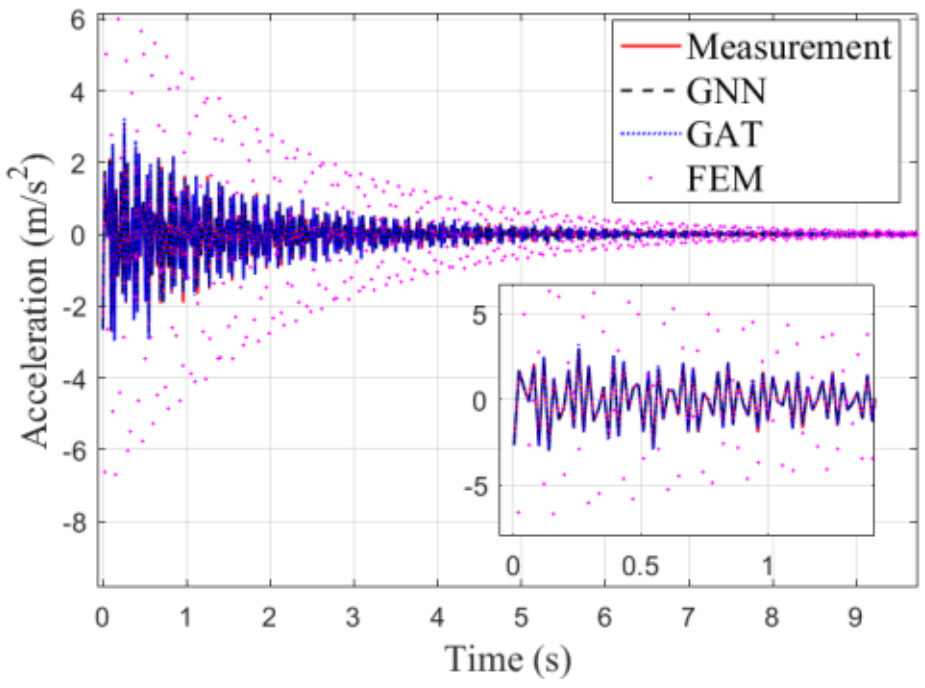

(*c*) Acc_3.

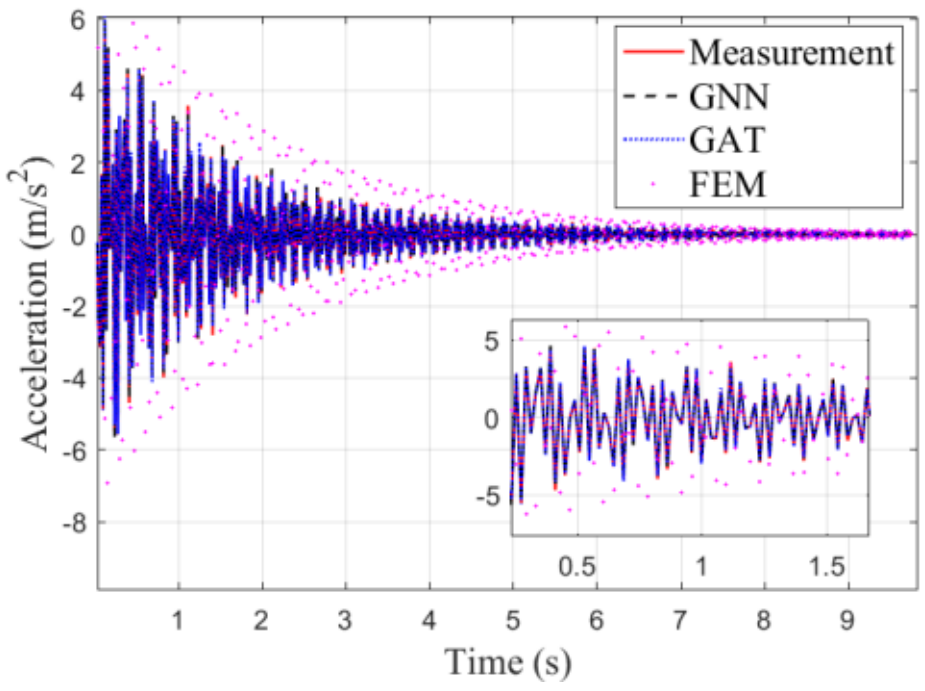

(*d*) Acc_4.

**Figure 12.** Partial time-domain acceleration response of 2S.

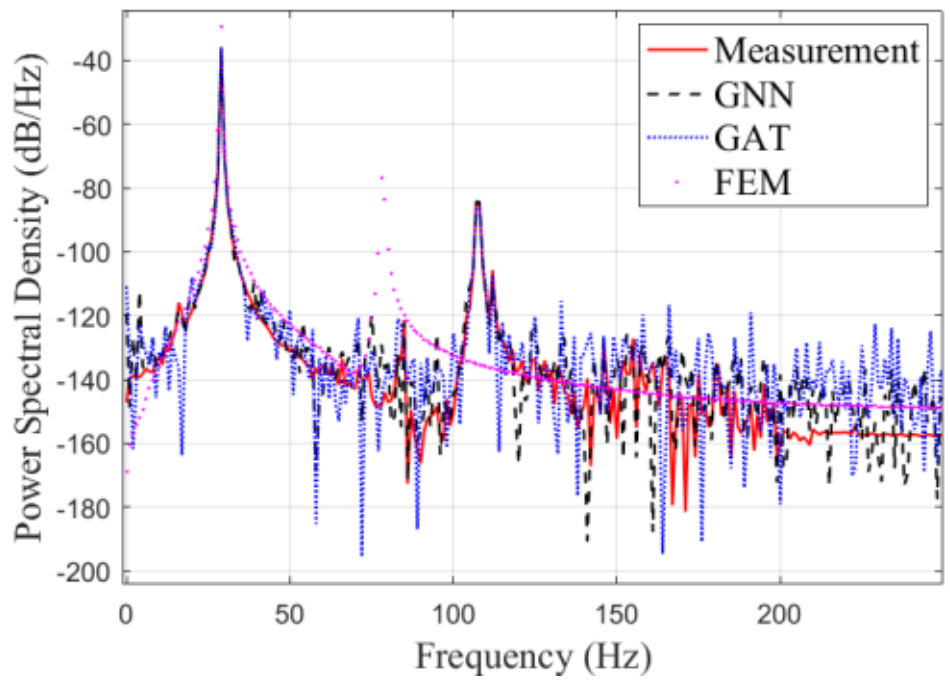

(*a*) Acc_1.

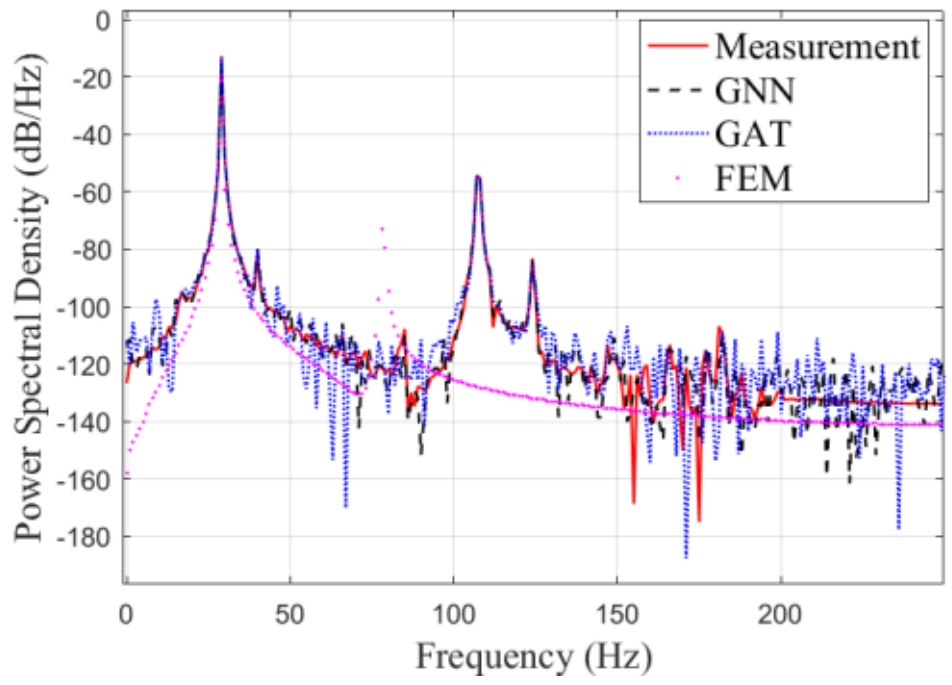

(*b*) Acc_2.

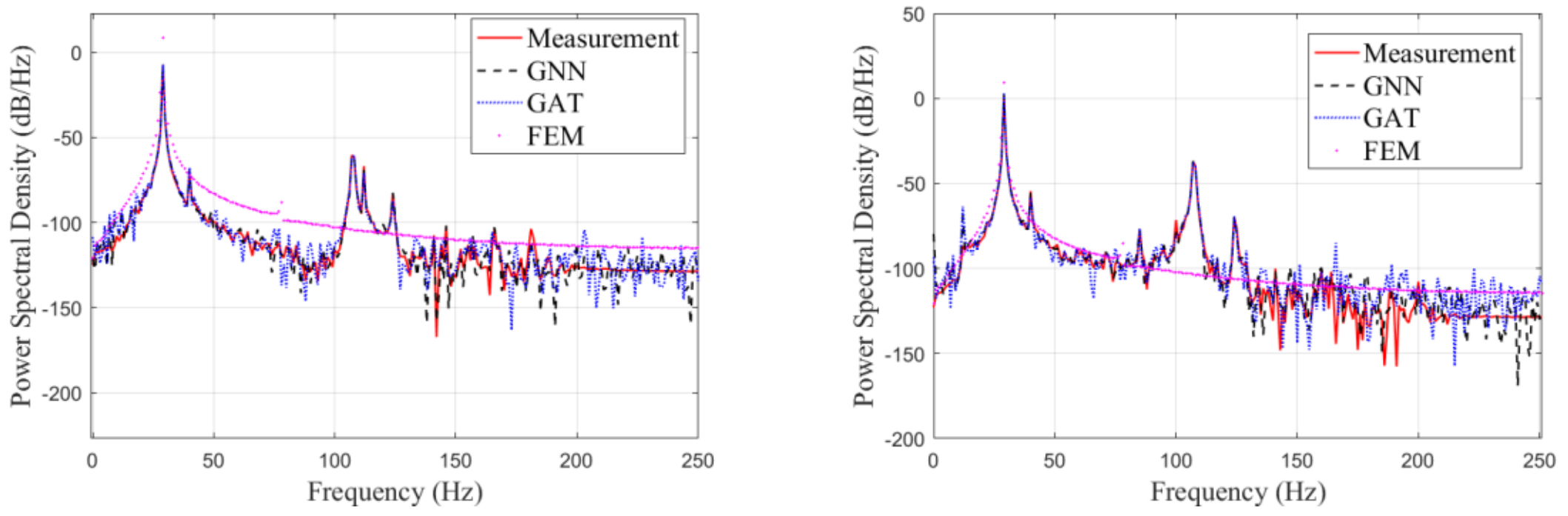

(*c*) Acc_3.    (*d*) Acc_4.

**Figure 13.** Partial acceleration response PSD of 2S structure.

Figure 12 clearly showed that the acceleration responses simulated by both the heterogeneous GNN and GAT methods closely matched the measured data. In contrast, the FEM numerical model displayed significant differences in both amplitude and phase compared to the measured values. The PSD analysis, Figure 13, further demonstrated that the GDTM methods closely corresponded to the measured data in the low-frequency range, though larger deviations were observed at higher frequencies. This indicated that the GDTM approaches more effectively predicted the experimental structural behavior, rather than capturing noise. Meanwhile, the FEM result exhibited considerable divergence from the measured values, even within the low-frequency range.

*4.2.3 Simulating dynamic response of 3S and 4S structure*

Additional experiments were carried out in this subsection. By modifying the dimensions of the adjacency matrix, the heterogeneous GNN-based and GAT-based models trained on 2S data was directly applied to simulate the structural responses of the 3S and 4S structures under impact loads. The results of these simulations were documented in Table 7.

As shown in Table 7, the accuracy of the acceleration responses simulated by both the heterogeneous GNN-based and GAT-based models for the 3S and 4S structures was largely consistent with the measured data. The overall vibration trends remained closely aligned, as evidenced by $R^2$ values consistently above 0.96. In contrast, the FEM simulation produced a poor match with the measured data, as indicated by the statistical metrics. Figures 14 and 15, as well as Figures 16 and 17, illustrated the acceleration time histories and PSD for the 3S and 4S structures, respectively.

**TABLE 7** Three evaluation metrics value of the Simulated Acceleration Responses of 3S and 4S.

| | **3S** | | | **4S** | | |
|---|---|---|---|---|---|---|
| | NMSE | $R^2$ | PE (%) | NMSE | $R^2$ | PE (%) |
| GNN | 0.0018 | 0.9981 | 9.55 | 0.0024 | 0.9935 | 12.12 |
| GAT | 0.0036 | 0.9884 | 15.39 | 0.034 | 0.9651 | 38.77 |
| FEM | 7.13 | 0.3318 | 224.64 | 6.45 | 0.5883 | 173.98 |

The simulation results of the acceleration responses for the 3S and 4S structures demonstrated that the heterogeneous GNN-based and GAT-based models effectively reproduced the structural behavior by merely adjusting the adjacency matrix. As depicted in Figures 14 and 16, both models generally captured the overall vibration patterns seen in the measured acceleration data. The PSD results, illustrated in Figures 15 and 17, further reinforced that the heterogeneous GDTM model's predictions corresponded closely with the measured data in the low-frequency. Nevertheless, the FEM generated responses showed the greatest divergence from the experimental data, with the PSD analysis indicating that FEM predictions only aligned with the measured accelerations at a few points in the low-frequency range.

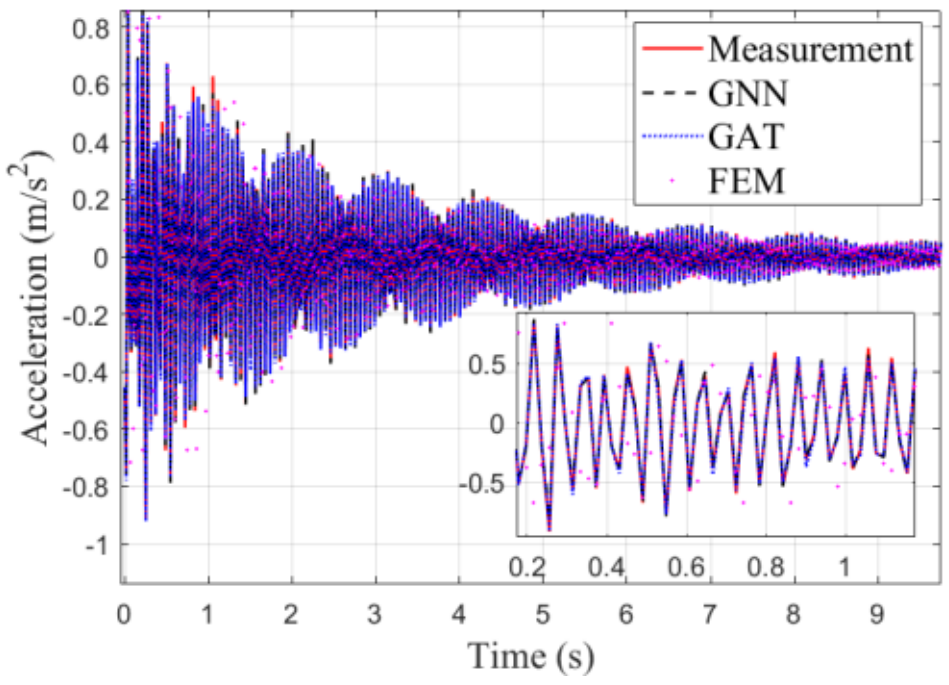

(*a*) Acc_2.

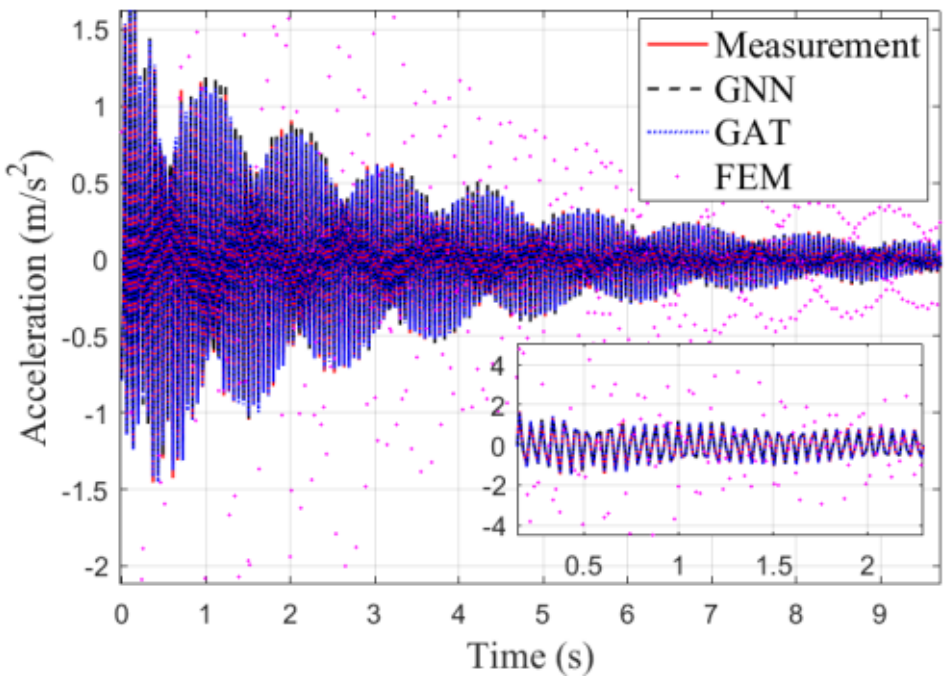

(*b*) Acc_4.

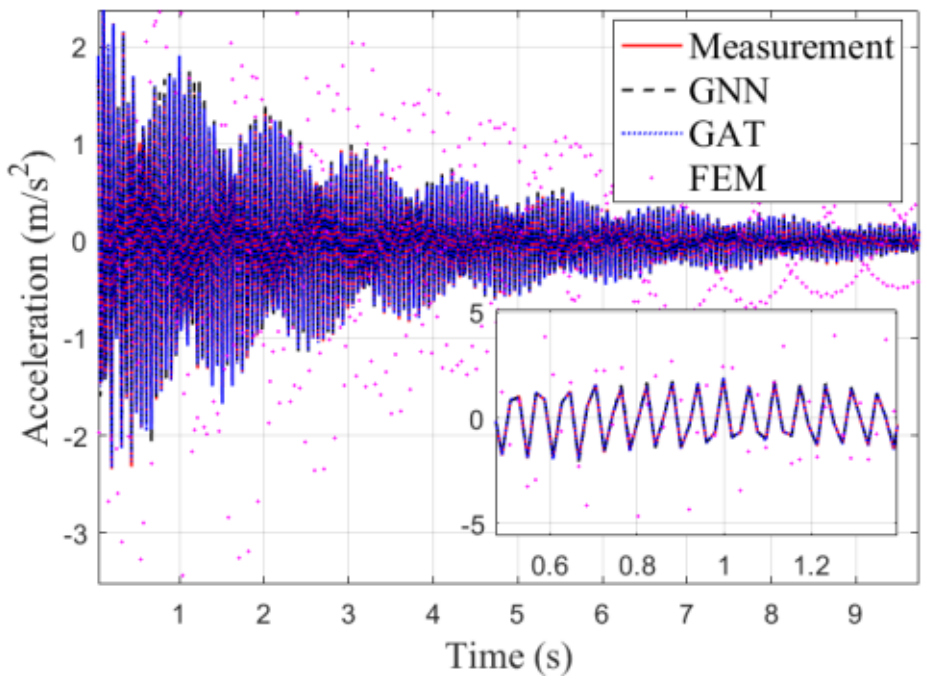

(*c*) Acc_6.

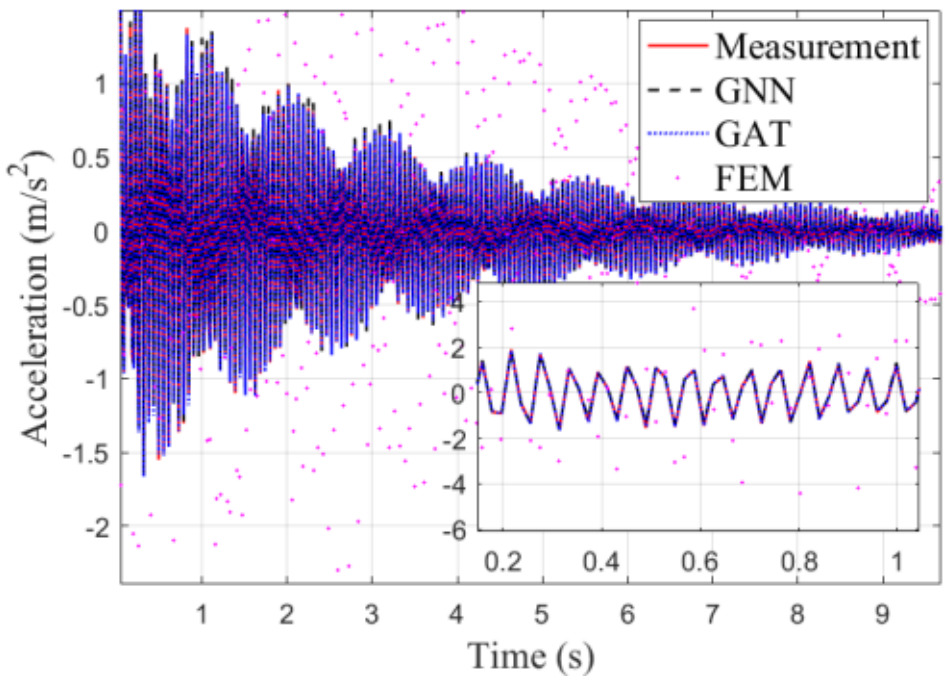

(*d*) Acc_8.

**Figure 14.** Partial time-domain acceleration response of 3S.

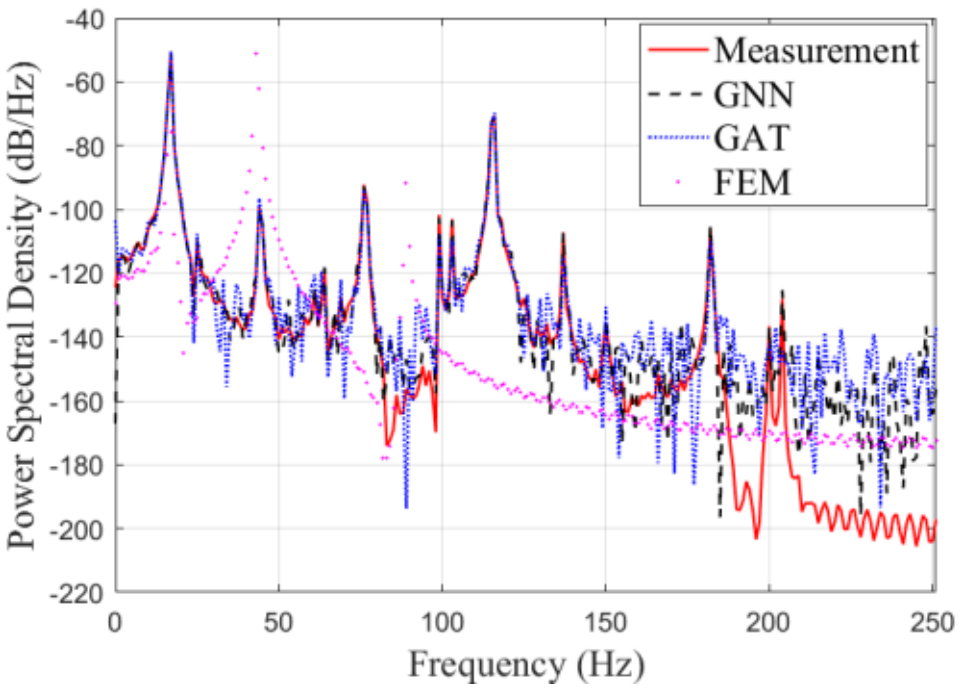

(*a*) Acc_2.

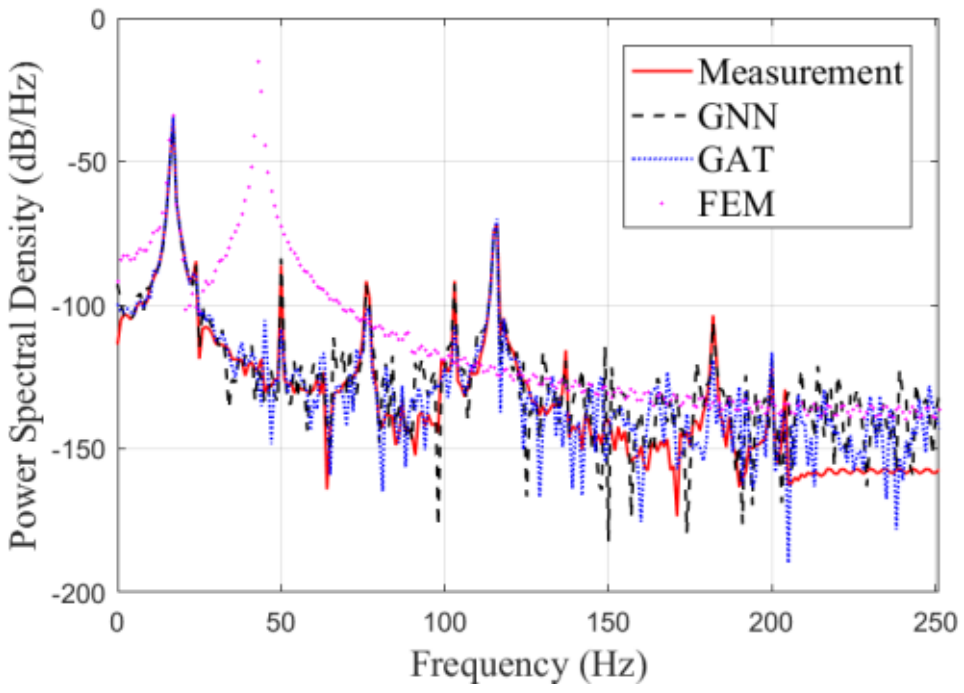

(*b*) Acc_4.

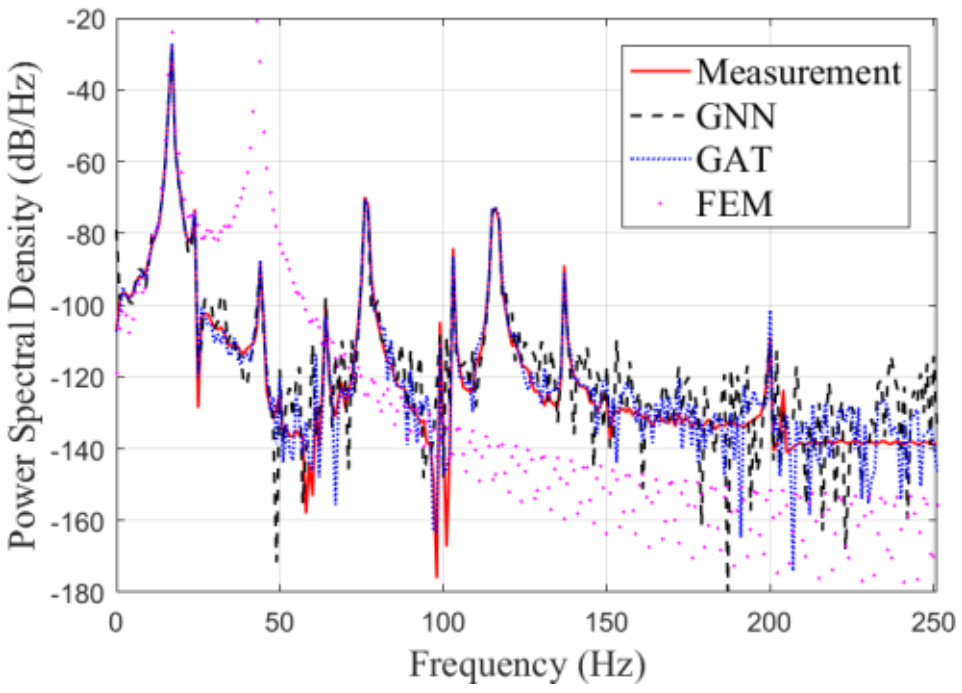

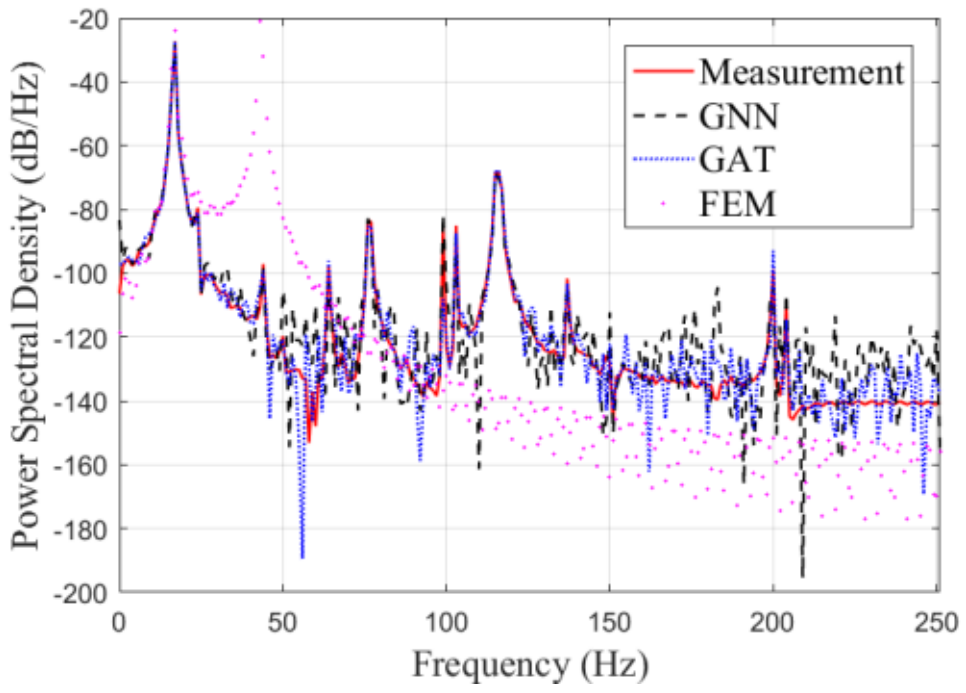

(*c*) Acc_6. (*d*) Acc_8.

**Figure 15.** Partial acceleration response PSD of 3S structure.

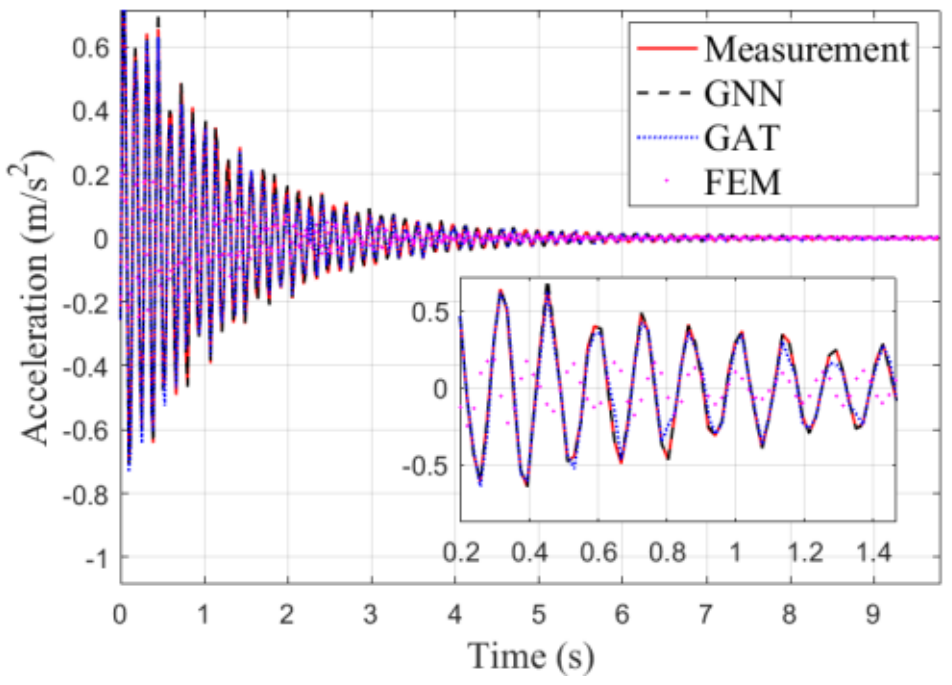

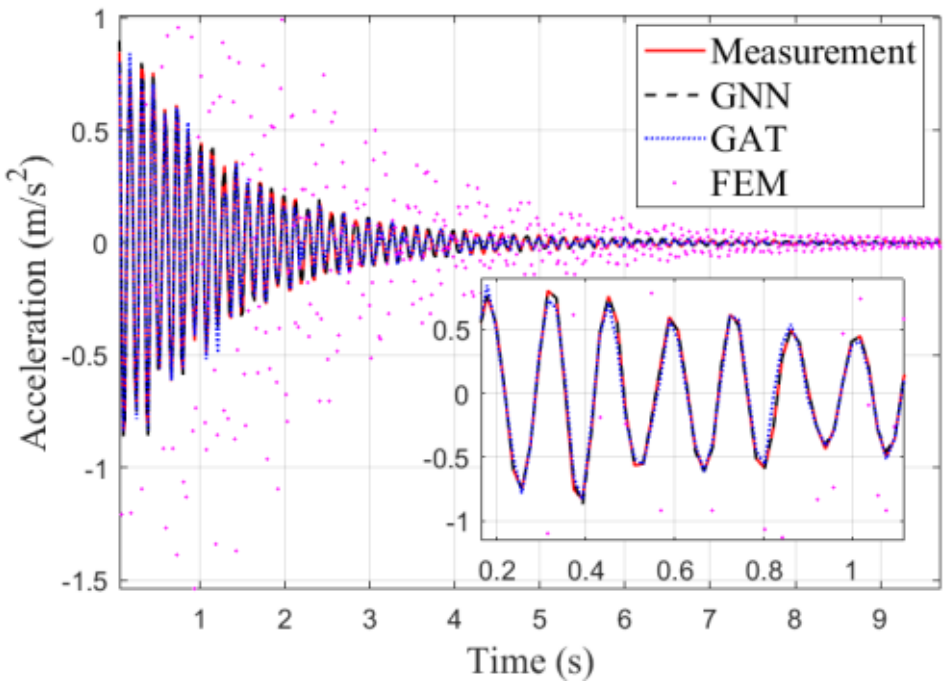

(*a*) Acc_2. (*b*) Acc_4.

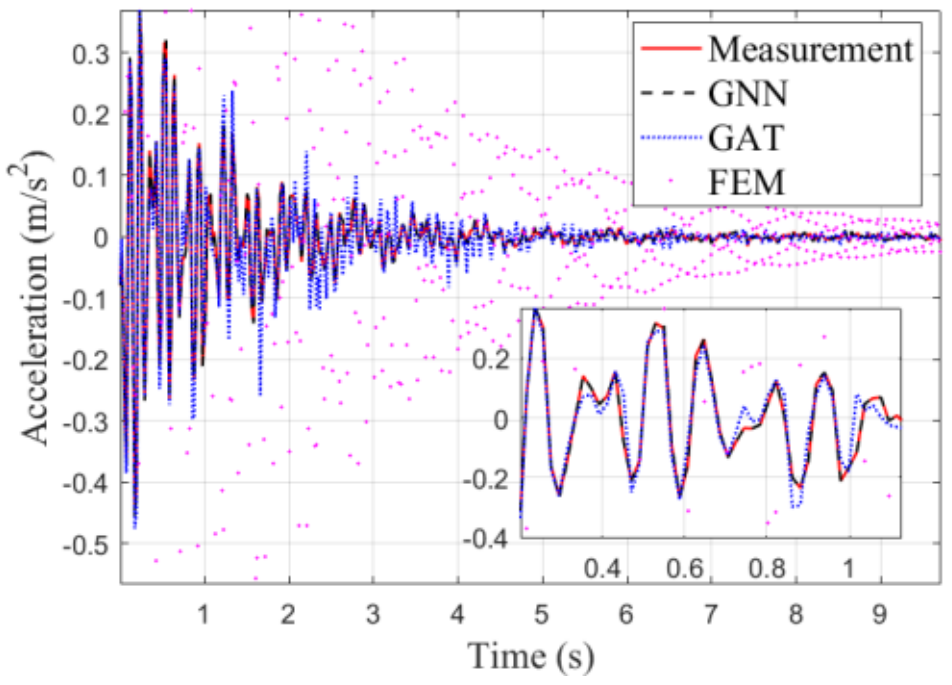

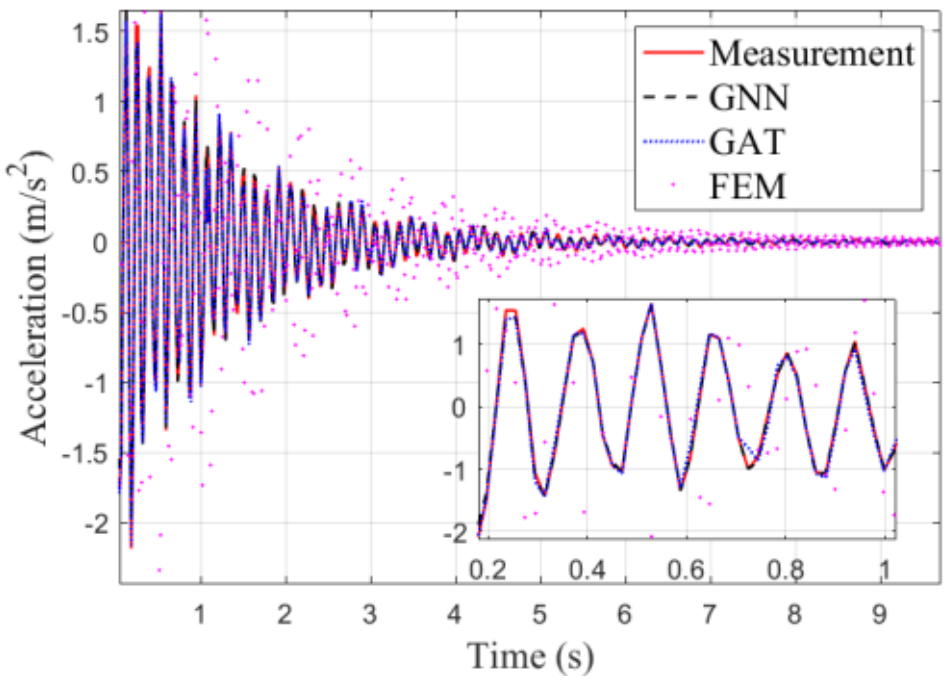

(*c*) Acc_6. (*d*) Acc_8.

**Figure 16.** Partial time-domain acceleration response of 4S

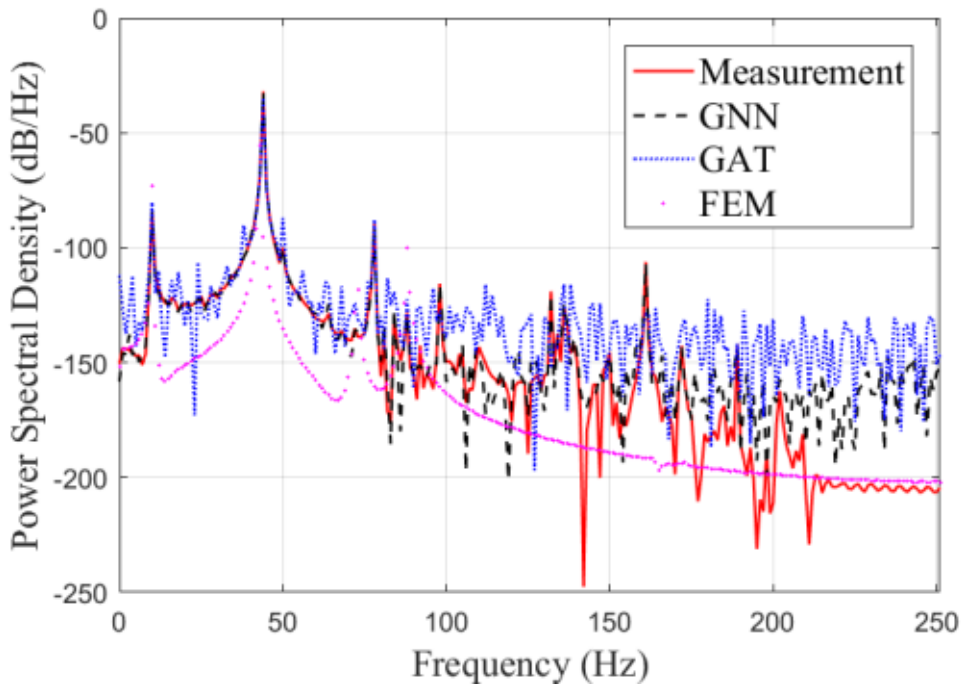

(*a*) Acc_2.

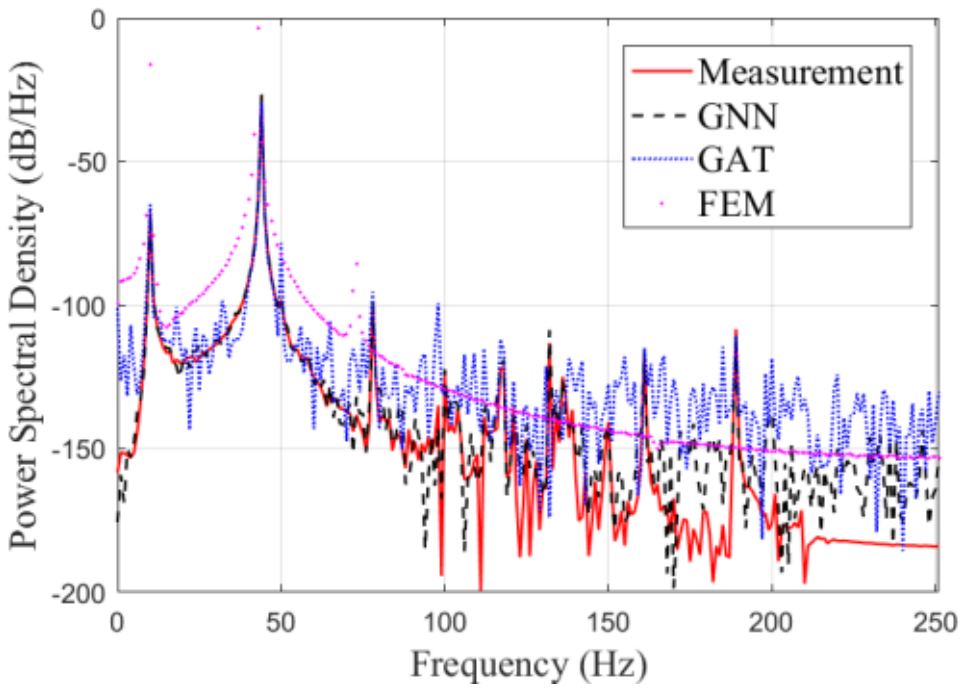

(*b*) Acc_4.

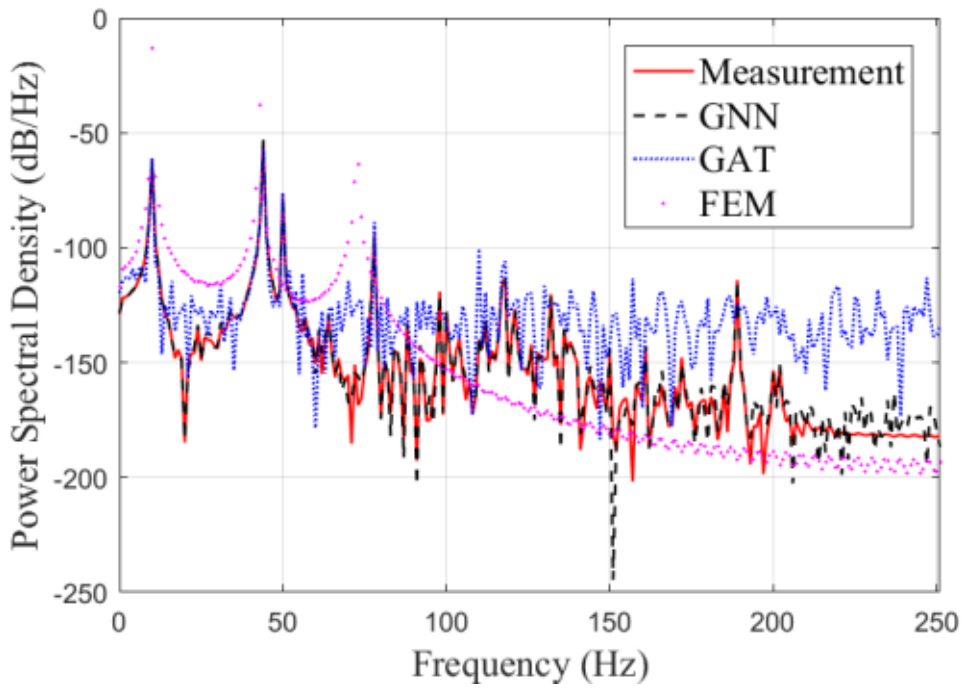

(*c*) Acc_6.

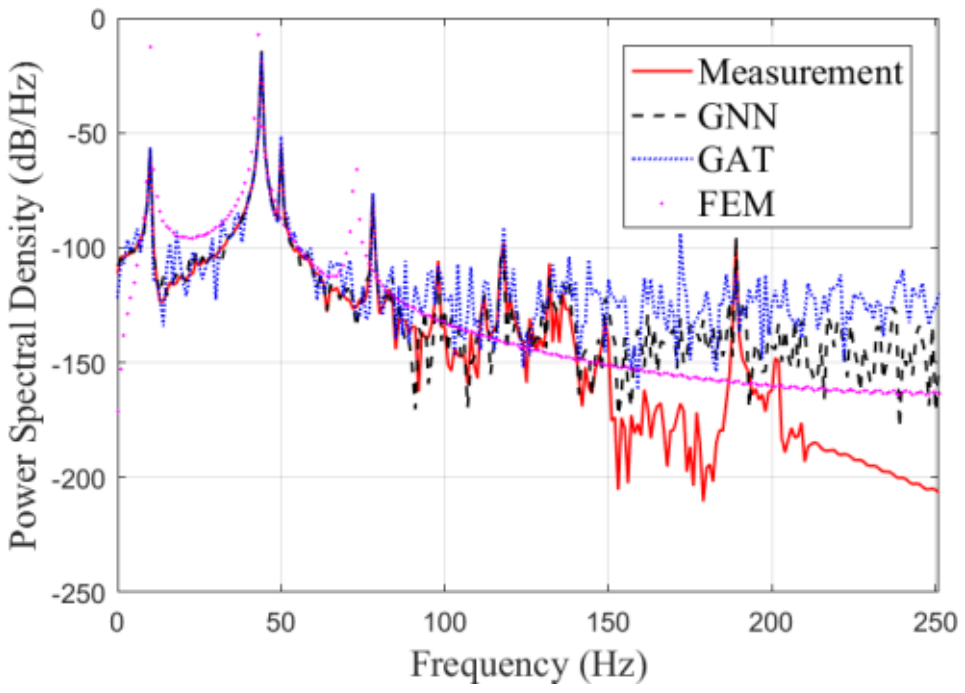

(*d*) Acc_8.

**Figure 17.** Partial acceleration response PSD of 4S structure.

#### *4.2.4 Discussion and efficiency comparison for the experimental case*

One of the key factors contributing to the inaccuracy of the FEM simulation was the use of translational and rotational springs to represent the behavior of IMCs during structural vibrations. The stiffness of these springs was derived through model updating techniques. During this process, the Artificial Ecosystem-Based Optimization (AEO) [58] algorithm was introduced to update these stiffness values by approximating the modal analysis (SSI-COV) results of the measured acceleration response [59]，as detailed in the Appendix, which is a common approach for simplifying IMC behavior in modular building structures[60-62]. However, numerous studies have highlighted the difficulty of accurately simulating the dynamic characteristics of IMCs, owing to their complex interaction mechanisms and the presence of contact nonlinearities[63-65]. Model updating based solely on frequency as the objective typically fails to reflect the true behavior of IMCs, especially their nonlinear aspects. Consequently, the FEM model's results were suboptimal. Given the excessive time requirements of transient analysis using FEM, time-domain model updating was not pursued in this study. Nevertheless, the comparison of the GDTM methods' simulation results with the measured acceleration responses demonstrated the effectiveness of the proposed approach. This breakthrough expanded the potential applications of deep learning in structural analysis and health monitoring.

**TABLE 8** Comparison of simulation efficiency.

| | Parameter quantity | Computing resources | Simulation time | | Parameter quantity | Computing resources | Simula tion |
|---|---|---|---|---|---|---|---|

| | | | | | | | | | time |
|---|---|---|---|---|---|---|---|---|---|
| 2S | GNN | 1751697 | **Heterogeneous GNN**: GPU(nvidia geforce rtx 3090 ti) | 19s | 3S | GNN | 1751697 | **GAT**: GPU(nvidia geforce rtx 3090 ti) | 19s |
| | GAT | 1755793 | | 19s | | GAT | 1755793 | | 19s |
| | FEM | Element:88 Node: 94 | | 1252s | | FEM | Element:132 Node: 142 | | 1524s |
| 4S | GNN | 1751697 | **FEM**: CPU(AMD Ryzen 7 5800H with Radeon Graphics) | 19s | | | | | |
| | GAT | 1755793 | | 19s | | | | | |
| | FEM | Element: 176 Node: 190 | | 1642s | | | | | |

In addition to achieving exceptional simulation accuracy, the two GDTM methods proposed in this study exhibited a substantial enhancement in computational efficiency compared to traditional numerical techniques. Table 8 detailed the parameter counts, computational resources, and the time required to simulate acceleration responses of equivalent duration for the FEM, heterogeneous GNN-based, and GAT-based models. The findings revealed that, with GPU assistance, the two GDTM methods realized a speed increase exceeding 80-fold in comparison to the FEM.

## *4.3 Interpretability of GDTM methods*

Clear model interpretability was the greatest advantage of the GDTM method over other deep learning algorithms. In this study, these were demonstrated through equations of discrete and continuous systems.

### *4.3.1 Interpretability of GDTM methods for discrete systems*

In section 4.1, Equation (2) described the motion equation of the MDOF system. In this case, **M**=M×**A**, **C**=C×**A**, **K**=K×**A** allowing Equation (2) to be reformulated as Equation (6). **A** represented the adjacency matrix used in the homogeneous GNN-based method. In Equation (6), the terms $\mathbf{A}\cdot\dot{\mathbf{Y}}(t)$ and $\mathbf{A}\cdot\mathbf{Y}(t)$ on the left side, and $\mathbf{L}\cdot\mathbf{F}(t)$ on the right side collectively formed the input data for the GDTM method, with $\ddot{\mathbf{Y}}(t)$ as the output. Consequently, training the GDTM model primarily involved fitting the system parameters $\mathrm{C}/M$, $\mathrm{K}/M$, and $M^{-1}$ , which provided the model with strong model interpretability.

$$\ddot{\mathbf{Y}}(t)+\mathrm{C}/M\cdot\mathbf{A}\cdot\dot{\mathbf{Y}}(t)+\mathrm{K}/M\cdot\mathbf{A}\cdot\mathbf{Y}(t)=M^{-1}\mathbf{LF}(t) \tag{6}$$

### *4.3.2 Interpretability of GDTM methods for continuous systems*

In section 4.2, the model's physical interpretation was elucidated by deriving it from the partial differential equations governing the vibrations of continuous systems. Elucidating its function within the computational process was key to grasping the underlying physical significance of the model. The principal vibration modes of continuous systems were categorized as tensile, torsional, shear, and bending vibrations. The partial differential equation governing the vibrations of such systems was derived from Equation (7) and Equation (8).

$$\frac{\partial^2\mathbf{u}}{\partial t^2}+2\varepsilon\cdot\frac{\partial\mathbf{u}}{\partial t}-c^2\cdot\frac{\partial^2\mathbf{u}}{\partial x^2}=\mathbf{P}(x,t) \tag{7}$$

In Equation (7), $\mathbf{u}(x, t)$ denoted the displacement response, which was defined as a function of both time $t$ and position $x$. The term $\mathbf{P}(x, t)$ represented the external excitation signal applied to the system. $\varepsilon=C(\mathrm{x})/2\rho A$, $C(\mathrm{x})$ described the damping coefficient per unit length, where $\rho$ and $A$ referred to the material's density and cross-sectional area, respectively. When $c^2=E/\rho$, with $E$ as the Young's modulus, Equation (7) expressed the tensile vibration of the continuous system. Alternatively, if $c^2=G/\rho$, where $G$ indicated the shear modulus, the equation described torsional vibration. Finally, if $c^2=\mu G/\rho$, where $\mu$ signified the sectional coefficient, Equation (7) captured the shear vibration of the system. Equation (8) represents the bending vibration of a continuous system (Bernoulli-Euler beam), where $K(\mathrm{x})= EJ/(\rho A)$, with $EJ$ denoting the flexural stiffness.

$$\frac{\partial^2 \mathbf{u}}{\partial t^2} + 2\varepsilon \cdot \frac{\partial \mathbf{u}}{\partial t} - K(x) \cdot \frac{\partial^4 \mathbf{u}}{\partial x^4} = \mathbf{P}(x,t) \tag{8}$$

The homogeneous general solutions to Equation (7) and (8) were presented in Equation (9) and (10), respectively. In Equation (9), $k=\omega/c$, where $\omega$ represented the system's natural frequency during free vibration. The time-dependent function $T(t)=Ae^{-\varepsilon t}\sin(\omega t+\varphi)$, with $A$, $\varphi$, $B_1$ and $B_2$ being constants derived from the system's initial conditions.

$$\mathbf{u}(x,t) = (B_1 \cdot sin\ kx + B_2 \cdot cos\ kx) \cdot T(t) \tag{9}$$

In Equation (10), $\mathrm{Y(x)}=L_1A_k(x)+L_2B_k(x)+L_3C_k(x)+L_4D_k(x)$, where $L_1$, $L_2$, $L_3$, and $L_4$ represented undetermined constants. $A_k(x)$, $B_k(x)$, $C_k(x)$, and $D_k(x)$ were Krylov functions, which could be derived from Equation (11) and (12).

$$\mathbf{u}(x,t) = Y(x) \cdot T(t) \tag{10}$$

$$\begin{cases} A_k(x) = \frac{ch\ kx + \cos kx}{2} \\ B_k(x) = \frac{sh\ kx + \sin kx}{2} \\ C_k(x) = \frac{ch\ kx - \cos kx}{2} \\ D_k(x) = \frac{sh\ kx - \sin kx}{2} \end{cases} \tag{11}$$

$$\begin{cases} ch\ kx = \frac{e^{kx} + e^{-kx}}{2} \\ sh\ kx = \frac{e^{kx} - e^{-kx}}{2} \\ cos\ kx = \frac{e^{ikx} + e^{-ikx}}{2} \\ sin\ kx = \frac{e^{ikx} - e^{-ikx}}{2} \end{cases} \tag{12}$$

In solving structural dynamic responses, the standard practice involved discretizing the dynamic equations within the time domain. Following this, the structural response at each discrete time step was iteratively calculated using direct integration methods, informed by the initial conditions. At the moment $t_0$, $T(t_0)$ reduced to a constant $T$. Consequently, Equation (9) was reformulated as Equation (13).

$$\begin{cases} \mathbf{u}_0 = (B_1 sin\ k \cdot 0 + B_2 cos\ k \cdot 0) \cdot T \\ \mathbf{u}_l = (B_1 sin\ k \cdot l + B_2 cos\ k \cdot l) \cdot T \end{cases} \tag{13}$$

Where, $\mathbf{u}_0$ and $\mathbf{u}_l$ represented the displacements at the two endpoints of a component. The relation $\mathbf{u}_l=G(\mathbf{u}_0)$ signified the displacement of a neighboring vertex when the vertex itself experienced a displacement of $\mathbf{u}_0$. From Equation (13), Equation (14) was derived. Similarly, Equation (15) was obtained from Equation (11).

$$\mathbf{u}_l = \left(B_1 \sin k \cdot l + \frac{\mathbf{u}_0}{T} \cos k \cdot l\right) \cdot T \tag{14}$$

$$\mathbf{u}_l = \left(\frac{\mathbf{u}_0}{T} A_k(l) + L_2 B_k(l) + L_3 C_k(l) + L_4 D_k(l)\right) \cdot T \tag{15}$$

Thus, the displacement $\mathbf{u}_i^1$ of any vertex $v_i$ at time $t_1$ could be expressed in terms of the displacements $\mathbf{u}_j^0$ of its neighboring vertices $v_j \in \mathbf{V}$ at time $t_0$ as follows: $\mathbf{u}_i^1 = f\,[\mathbf{u}_i^0, G_1(\mathbf{u}_{j1}^0), \cdots, G_n(\mathbf{u}_{jn}^0), \mathbf{P}(v_i, t_1), \Delta t]$. In contrast, the GDTM models represented this relationship as: $\mathbf{u}_i^1 = f\,\{\sigma[\boldsymbol{w}(\mathbf{A}_i \times \mathbf{U}^0 \parallel \mathbf{A}_{j1} \times \mathbf{U}^0 \parallel \cdots \parallel \mathbf{A}_{jn} \times \mathbf{U}^0 \parallel \mathbf{P}) + \boldsymbol{b}], \Delta t\}$. In this context, $f(\cdot)$ denoted the update of vertex features, while $\sigma[\boldsymbol{w}(\cdot)+\boldsymbol{b}]$ encapsulated the weights, biases, and nonlinear activation function of the MLP. The term $\Delta t$ signified the integration time interval. $\mathbf{U}^0$ represented the displacement matrix for all vertices at time $t_0$ , and $\mathbf{u}_i$ was articulated as the product of the $i$-th row vector $\mathbf{A}_i$ from the adjacency matrix and the displacement matrix $\mathbf{U}$. The symbol $\parallel$ represented the concatenation operation. Consequently, it became apparent that the MLP within the GDTM model realized two function: ① the function $G(\cdot)$, as illustrated in Equation (14) and (15). ② the $f(\cdot)$, which facilitated the update of vertex features.

Furthermore, the attention coefficients (AC) updated at each time step by the GAT-based model were recorded to clarify their physical interpretation. Meanwhile, Figure 18 illustrated the time-varying curves of select attention coefficients. Among them, $AC_c$, $AC_{IMC}$, and $AC_b$ represented the attention coefficients of a column, an IMC, and a beam, respectively. At the same time, the attention coefficients of the column in different directions, clockwise and anticlockwise, were compared.

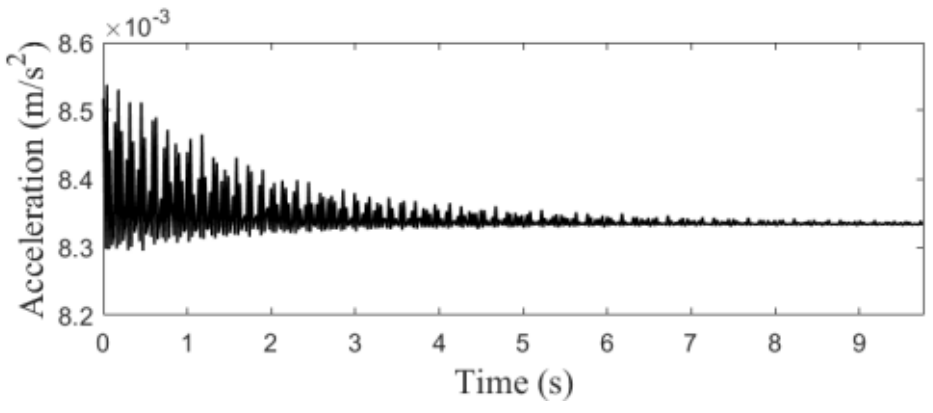

(*a*) $AC_c$ (clockwise)

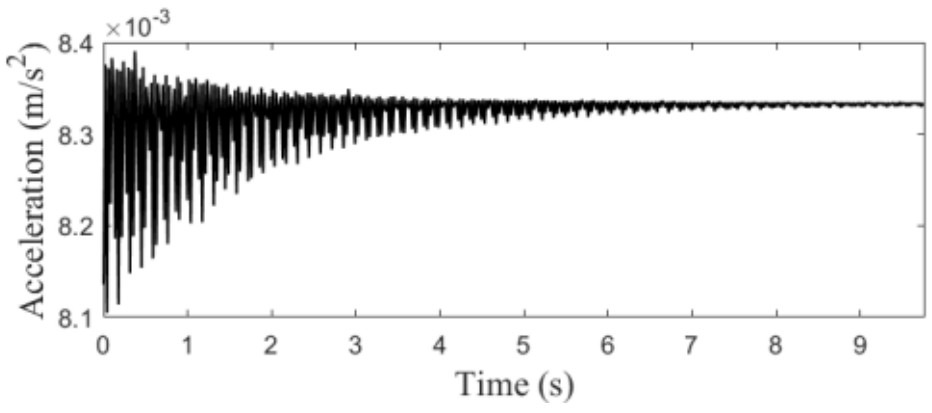

(*b*) $AC_c$ (anticlockwise)

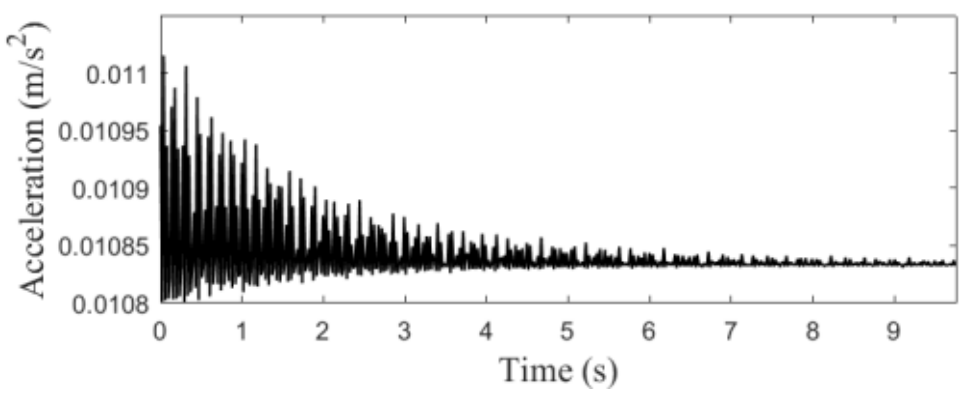

(*c*) $AC_{IMC}$

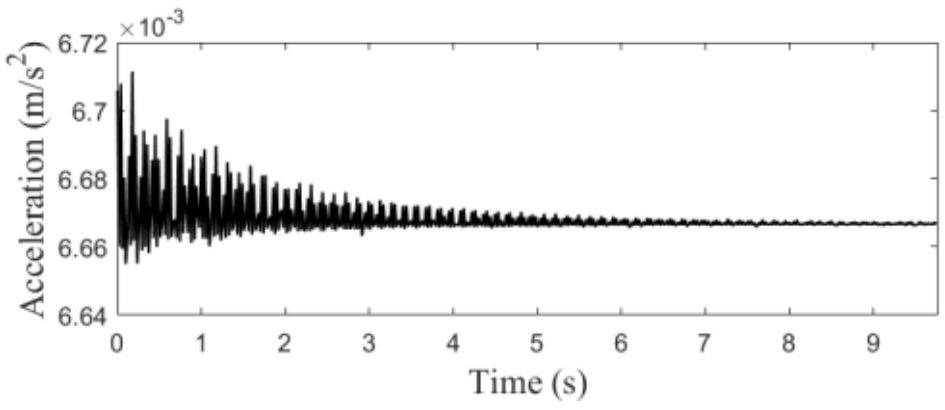

(*d*) $AC_b$

**Figure 18.** The attention coefficients of GAT-based method.

As depicted in Figure 18, although the softmax operation caused all attention coefficients to remain greater than zero, it was still clear that the fluctuation patterns of $AC_c$ (clockwise) and $AC_c$ (anticlockwise) moved in opposite directions. This demonstrated that attention coefficients in opposite directions along the same edge behaved inversely, consistent with the principle of action and reaction forces. Furthermore, the variations in $AC_c$, $AC_{IMC}$, and $AC_b$ revealed that the three attention coefficients corresponding to different components eventually stabilized at distinct values, highlighting the differences between the properties of the column, IMC, and beam. In addition, although the fluctuation directions differed, the amplitude of all attention coefficients progressively diminished over time, which aligned with the energy dissipation effect of damping in the structure.

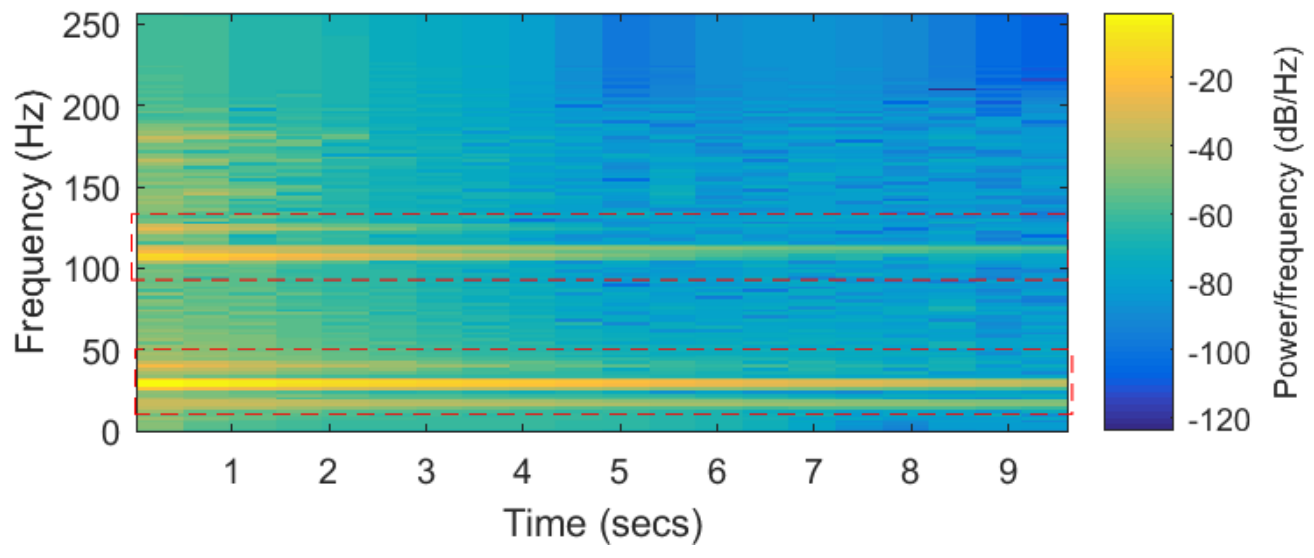

(*a*) STFT of acceleration from Acc_3 position of 2S(Measured)

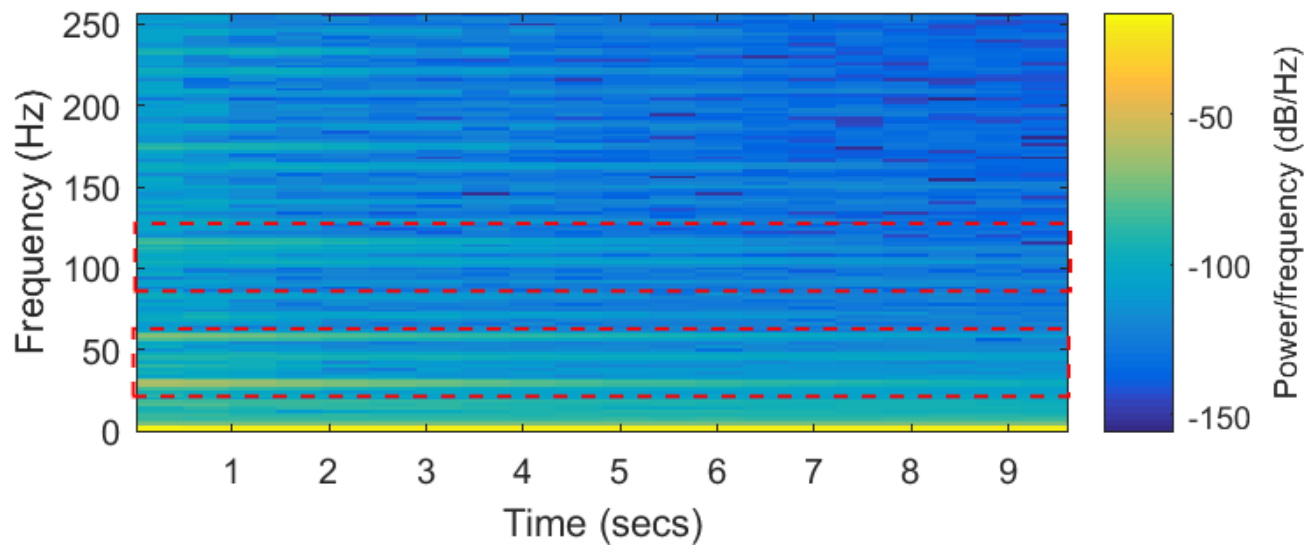

(*b*) STFT of attention coefficients of an IMC of 2S

**Figure 19.** STFT analysis results of 2S.

The attention coefficients of an IMC component in 2S structure were examined using the Short-Time Fourier Transform (STFT) and compared against the measured acceleration response show as Figure 19. In Figure 19, both exhibited highlight regions in the frequency bands of (0 Hz, 50 Hz) and (100 Hz, 150 Hz), with these values progressively diminishing over time. This behavior was attributed to the structure's nonlinear characteristics. The findings demonstrated a clear time-frequency alignment between the attention coefficients and the measured acceleration response.

*4.3.3 Discussion*

This section demonstrated the interpretability of the GDTM model by deriving vibration equations for both discrete and continuous systems, clarifying the roles of its two core components—the adjacency matrix and the MLP—in modeling vibration transmission and aggregation across vertices. Viewed from a symbolic perspective—a major AI approach alongside neural networks—the GDTM model combined automated data and pattern feature extraction with manually encoded rules. This approach enhanced the model's ability to represent complex physical laws and unknown parameters, while also expanding the applicability of deep neural networks to a range of structured tasks. Improved interpretability greatly increased the reliability of neural network models.

## 5. Conclusion

To better support structural dynamics analysis and health monitoring, this study proposed three GDTM methods that significantly enhance the accuracy and efficiency of simulations. Key findings

are as follows:

(1) Numerical simulations using a 10-DOF system demonstrated the model's accuracy in capturing dynamic responses under various loads, with NMSE as low as $5.02\times10{-}3$. Adjusting the adjacency matrix enabled accurate responses for systems with diverse topologies and parameters, achieving NMSE no more than 0.013 for the 5-DOF, 12-DOF, 20-DOF, and 30-DOF systems.

(2) In the experimental cases, three planar modular steel frameworks were constructed to validate the GDTM approach. Trained on the 2S structure, the heterogeneous GNN- and GAT-based models effectively captured vertex acceleration across all frameworks. For the 2S structure, the heterogeneous GNN-based model achieved an NMSE of 0.0015, while the GAT-based model recorded 0.0038. In the 3S structure, the NMSE was 0.0018 for the heterogeneous GNN-based model and 0.0036 for the GAT-based model. For the 4S structure, the heterogeneous GNN-based model achieved an NMSE of 0.0024, while the GAT-based model showed a larger deviation of 0.034. Compared to FEM, the GDTM approach significantly improved simulation accuracy and increased computational efficiency by over 80 times.

(4) The interpretability of GDTM was illustrated by deriving vibration equations and explaining the adjacency matrix and MLP. GDTM accurately simulated dynamic responses in discrete and continuous systems, ensuring reliability across diverse structural topologies.

In conclusion, GDTM offers improved accuracy, computational speed, and physical interpretability, making it a promising tool for structural dynamics and health monitoring.

## Declaration of conflicting interests

The author(s) declared no potential conflicts of interest with respect to the research, authorship, and/or publication of this article.

## Data Availability

The data that support the findings of this study are available on request from the corresponding author. The data are not publicly available due to privacy or ethical restrictions.

## Conflicts of Interest

The authors declare that they have no conflict of interest.

## Authors' Contributions

Jun Zhang and Tong Zhang contributed equally to this work.

## Acknowledgments

The authors gratefully acknowledge the financial support provided by the National Key Research and Development Program of China (Grant No. 2022YFB2602103).

## References

[1] Wei, X. C., Fan, J. S., Liu, Y. F., Zhang, J. X., Liu, X. G., & Kong, S. Y. (2022). Automated inspection and monitoring of member deformation in grid structures. *Computer-Aided Civil and Infrastructure Engineering, 37*(10), 1277-1297.

[2] Xu, M. N., Sun, L. M., Liu, Y. F., Li, B. L., Zhou, M., & Chen, X. (2024). Member separation and deformation recognition of spatial grid structures in-service. *Engineering Structures*, *304*, 117642.

[3] Zhang, A., Ma, H., Zhao, X., Zhang, Y., Wang, J., & Su, M. (2024). 3D laser scanning for automated structural modeling and deviation monitoring of multi-section prefabricated cable domes. *Automation in Construction*, *165*, 105573.

[4] Ay, A. M., & Wang, Y. (2014). Structural damage identification based on self-fitting armax model and multi-sensor data fusion. *Structural Health Monitoring*, *13*(4), 445-460.

[5] Wang, Y., & Hao, H. (2010). An introduction to compressive sensing and its potential applications in structural engineering. *In The 11th international symposium on structural engineering* (p. 1089-1094).

[6] Zhou, J., Kato, B., & Wang, Y. (2023). Operational modal analysis with compressed measurements based on prior information. *Measurement, 211*, 112644.

[7] Zhou., J., Yu., S., Li., H., Wang., Y., & Ou., J. (2023). Automated operational modal analysis for civil engineering structures with compressed measurements. *Measurement, 29*(223), 11377.

[8] Wang, Y., & Hao, H. (2015). Damage identification scheme based on compressive sensing. *Journal of Computing in Civil Engineering, 29*(2), 04014037.

[9] Zhang, T., Biswal, S., & Wang, Y. (2020). Shmnet: Condition assessment of bolted connection with beyond human-level performance. *Structural Health Monitoring, 19*(4), 1188-1201.

[10] Zhang, T., & Wang, Y. (2019). Deep learning algorithms for structural condition identification with limited monitoring data. In *In international conference on smart infrastructure and construction 2019 (icsic) driving data-informed decision-making* (p. 421-426). London, UK.

[11] Bao, N., Zhang, T., Huang, R., S, B., Su, J., & Wang, Y. (2023). A deep transfer learning network for structural condition identification with limited real-world training data. *Structural Control and Health Monitoring*, *2023*(1), 8899806.

[12] Lin, Y. Z., Nie, Z. H., & Ma, H. W. (2022). Dynamics-based cross-domain structural damage detection through deep transfer learning. *Computer-Aided Civil and Infrastructure Engineering*, *37*(1), 24-54.

[13] Mahbubi Motlagh, N., & Noorzad, A. (2021). Discrete element method simulation of dynamic behavior of granular materials. *Amirkabir Journal of Civil Engineering, 53*(10), 4325-4344.

[14] Wang, T., Zhang, F., Furtney, J., & Damjanac, B. (2022). A review of methods, applications and limitations for incorporating fluid flow in the discrete element method. *Journal of Rock Mechanics and Geotechnical Engineering, 14*(3), 1005-1024.

[15] Kim, H., Wagoner, M. P., & Buttlar, W. G. (2008). Simulation of fracture behavior in asphalt concrete using a heterogeneous cohesive zone discrete element model. *Journal of materials in civil engineering, 20*(8), 552-563.

[16] Marina, S., Derek, I., Mohamed, P., Yong, S., & Imo-Imo, E. K. (2015). Simulation of the hydraulic fracturing process of fractured rocks by the discrete element method. *Environmental earth*

*sciences, 73*, 8451-8469.
[17] Beskos, D. E. (1995). Dynamic inelastic structural analysis by boundary element methods. *Archives of Computational Methods in Engineering*, *2*, 55-87.
[18] Newman, J. N., & Lee, C. H. (2002). Boundary-element methods in offshore structure analysis. *Journal Of Offshore Mechanics And Arctic Engineering, 124*(2), 81-89.
[19] Wang, Y., & Hao, H. (2014). Modelling of guided wave propagation with spectral element: application in structural engineering. *Applied Mechanics and Materials, 553*, 687-692.
[20] Wang, Y., Hao, H., Zhu, X., & Ou, J. (2012). Spectral element modelling of wave propagation with boundary and structural discontinuity reflections. *Advances in Structural Engineering, 15*(5), 855-870.
[21] Wang, Y., Zhu, X., Hao, H., & Ou, J. (2009). Guided wave propagation and spectral element method for debonding damage assessment in rc structures. *Journal of sound and vibration, 324*((3-5)), 751-772.
[22] Wang, Y., Zhu, X., Hao, H., & Ou, J. (2011). Spectral element model updating for damage identification using clonal selection algorithm. *Advances in Structural Engineering, 14*(5), 837-856.
[23] Fancher, S., Purohit, P., & Katifori, E. (2023). An efficient spectral method for the dynamic behavior of truss structures. *arXiv preprint arXiv:2309.02448*.
[24] Lee, U. (2009). *Spectral element method in structural dynamics*. John Wiley & Sons.
[25] Ma, R., Xia, J., Chang, H., Xu, B., & Zhang, L. (2021). Experimental and numerical investigation of mechanical properties on novel modular connections with superimposed beams. *Engineering Structures*, *232*, 111858.
[26] Fu, Y., Liang, J., Wang, Y., & Ou, J. (2025). Updating complex boundary conditions using latent system internal forces towards structural digital twin models. *Mechanical Systems and Signal Processing*, *224*, 112088.
[27] Zhan, J., Zhang, F., Siahkouhi, M., Kong, X., & Xia, H. (2021). A damage identification method for connections of adjacent box-beam bridges using vehicle–bridge interaction analysis and model updating. *Engineering Structures*, *228*, 111551.
[28] Liang, J., Kato, B., & Wang, Y. (2023). Constructing simplified models for dynamic analysis of monopile-supported offshore wind turbines. *Ocean Engineering, 271*, 113785.
[29] Ni, Y. Q., Xia, Y., Lin, W., Chen, W. H., & Ko, J. M. (2012). Shm benchmark for high-rise structures: a reduced-order finite element model and field measurement data. *Smart Structures and Systems, 10*(4), 411-426.
[30] Ren, X., Xu, Y., Shen, T., Wang, Y., & Bhattacharya, S. (2023). Support condition monitoring of monopile-supported offshore wind turbines in layered soil based on model updating. *Marine Structures, 87*, 103342.
[31] Nayek, R., Chakraborty, S., & Narasimhan, S. (2019). A gaussian process latent force model for joint input-state estimation in linear structural systems. *Mechanical Systems and Signal Processing, 128*, 497-530.
[32] Rogers, T. J., Worden, K., & Cross, E. J. (2020). Bayesian joint input-state estimation for nonlinear systems. Vibration, 3(3), 281-303.
[33] Biswal, S., Chryssanthopoulos, M. K., & Wang, Y. (2022). Condition identification of bolted connections using a virtual viscous damper. *Structural health monitoring, 21*(2), 731-752.
[34] Fu, Y., & Wang, Y. (2023). Updating numerical models towards time domain alignment of structural dynamic responses with a limited number of sensors. *Mechanical Systems and Signal*

*Processing, 204*, 110759.
[35] Kutz, J. N., Brunton, S. L., Brunton, B. W., & Proctor, J. L. (2016). Dynamic mode decomposition: data-driven modeling of complex systems. *Society for Industrial and Applied Mathematics*.
[36] Moore, B. (1981). Principal component analysis in linear systems: Controllability, observability, and model reduction. *IEEE transactions on automatic control, 26*(1), 17-32.
[37] Cabell, R. H., & Fuller, C. R. (1999). A principal component algorithm for feedforward active noise and vibration control. *Journal of Sound and Vibration, 227*(1), 159-181.
[38] Luo, H., & Paal, S. G. (2023). A data-free, support vector machine-based physics-driven estimator for dynamic response computation. *Computer-Aided Civil and Infrastructure Engineering, 38*(1), 26-48.
[39] Yinfeng, D., Yingmin, L., Ming, L., & Mingkui, X. (2008). Nonlinear structural response prediction based on support vector machines. *Journal of Sound and Vibration, 311*(3-5), 886-897.
[40] Cybenko, G. (1989). Approximation by superpositions of a sigmoidal function. *Mathematics of control, signals and systems, 2*(4), 303-314.
[41] Huang, C. S., Hung, S. L., Wen, C. M., & Tu, T. T. (2003). A neural network approach for structural identification and diagnosis of a building from seismic response data. *Earthquake engineering structural dynamics, 32*(2), 187-206.
[42] Lagaros, N. D., & Papadrakakis, M. (2012). Neural network based prediction schemes of the non-linear seismic response of 3d buildings. *Advances in Engineering Software, 44*(1), 92-115.
[43] Oh, B. K., Park, Y., & Park, H. S. (2020). Seismic response prediction method for building structures using convolutional neural network. *Structural Control and Health Monitoring, 27*(5), e2519.
[44] Xue, J., Xiang, Z., & Ou, G. (2021). Predicting single freestanding transmission tower time history response during complex wind input through a convolutional neural network based surrogate model. *Engineering Structures, 233*, 111859.
[45] Torky, A. A., & Ohno, S. (2021). Deep learning techniques for predicting nonlinear multi-component seismic responses of structural buildings. Computers Structures, 252, 106570.
[46] Xu, Z., Chen, J., Shen, J., & Xiang, M. (2022). Recursive long short-term memory network for predicting nonlinear structural seismic response. *Engineering Structures, 250*, 113406.
[47] Yun, D. Y., & Park, H. S. (2024). Noise-robust structural response estimation method using short-time Fourier transform and long short-term memory. *Computer-Aided Civil and Infrastructure Engineering*.
[48] Zhang, C., Tao, M. X., Wang, C., & Fan, J. S. (2024). End-to-end generation of structural topology for complex architectural layouts with graph neural networks. *Computer-Aided Civil and Infrastructure Engineering*, *39*(5), 756-775.
[49] Kipf, Thomas N., & Welling, Max. (2017). Semi-supervised classification with graph convolutional networks. *International Conference on Learning Representations* (ICLR).
[50] Veličković, P., Cucurull, G., Casanova, A., Romero, A., Lio, P., & Bengio, Y. (2018). Graph attention networks. *International Conference on Learning Representations* (ICLR).
[51] Hamilton, W., Ying, Z., & Leskovec, J. (2017). Inductive representation learning on large graphs. *Advances in neural information processing systems, 30*.
[52] Battaglia, P., Pascanu, R., Lai, M., & Jimenez Rezende, D. (2016). Interaction networks for learning about objects, relations and physics. *Advances in neural information processing*

*systems*, *29*.
[53] Sanchez-Gonzalez, A., Godwin, J., Pfaff, T., Ying, R., Leskovec, J., & Battaglia, P. (2020, November). Learning to simulate complex physics with graph networks. In *International conference on machine learning* (pp. 8459-8468). PMLR.
[54] Zhang, J., Zhang, T., & Wang, Y. (2022). Gnn-based structural dynamics simulation for modular buildings. In *In Chinese conference on pattern recognition and computer vision (prcv)* (p. 245-258). China, Shenzhen.
[55] Song, L. H., Wang, C., Fan, J. S., & Lu, H. M. (2023). Elastic structural analysis based on graph neural network without labeled data. *Computer-Aided Civil and Infrastructure Engineering, 38*(10), 1307-1323.
[56] Li, Q., Wang, Z., Li, L., Hao, H., Chen, W., & Shao, Y. (2023). Machine learning prediction of structural dynamic responses using graph neural networks. *Computers Structures, 289*, 107188.
[57] Kuo, P. C., Chou, Y. T., Li, K. Y., Chang, W. T., Huang, Y. N., & Chen, C. S. (2024). Gnn-lstm-based fusion model for structural dynamic responses prediction. *Engineering Structures, 306*, 117733.
[58] Peeters, B., & De Roeck, G. (1999). Reference-based stochastic subspace identification for output-only modal analysis. *Mechanical systems and signal processing, 13*(6), 855-878.
[59] Zhao, W., Wang, L., & Zhang, Z. (2020). Artificial ecosystem-based optimization: a novel nature-inspired meta-heuristic algorithm. *Neural Computing and Applications, 32*(13), 9383-9425.
[60] Lacey, A. W., Chen, W., Hao, H., & Bi, K. (2020). Effect of inter-module connection stiffness on structural response of a modular steel building subjected to wind and earthquake load. *Engineering Structures*, *213*, 110628.
[61] Lacey, A. W., Chen, W., Hao, H., & Bi, K. (2021). Lateral behaviour of modular steel building with simplified models of new inter-module connections. *Engineering Structures*, *236*, 112103.
[62] CEN. EN 1993-1-8:2005 Eurocode 3: Design of steel structures - Part 1-8: Design of joints. Brussels, Belgium: European Committee for Standardization (CEN); 2005.
[63] Styles, A. J., Luo, F. J., Bai, Y., & Murray-Parkes, J. B. (2016). Effects of joint rotational stiffness on structural responses of multi-story modular buildings. In *Transforming the Future of Infrastructure through Smarter Information: Proceedings of the International Conference on Smart Infrastructure and ConstructionConstruction, 27–29 June 2016* (pp. 457-462). ICE publishing.
[64] Lacey, A. W., Chen, W., Hao, H., & Bi, K. (2019). Review of bolted inter-module connections in modular steel buildings. *Journal of Building Engineering*, *23*, 207-219.
[65] Chan, T. M., & Chung, K. F. (2021). Effect of inter-module connections on progressive collapse behaviour of MiC structures. *Journal of Constructional Steel Research*, *185*, 106823.

## Appendix

Numerical models grounded in FEM principles were developed using Ansys software, and their results were juxtaposed against those generated from the GDTM methodology. The experimental structures encompassed two-story (2S), three-story (3S), and four-story (4S) structures, comprising columns, beams, and inter-module connections (IMCs). To represent the columns and beams, Beam188 elements were employed, while Combin14 elements, incorporating spring-dampers that captured both translational and rotational stiffness, were utilized to simulate the IMCs as well as the boundary conditions at the structural base. The steel materials were defined by a density of $7.85 \times 10^3$ kg/m³, a Poisson's ratio of 0.25, and a Young's modulus of 206 GPa. Figure 20 presents the constructed model, meshing characteristics, and the applied connectivity constraints.

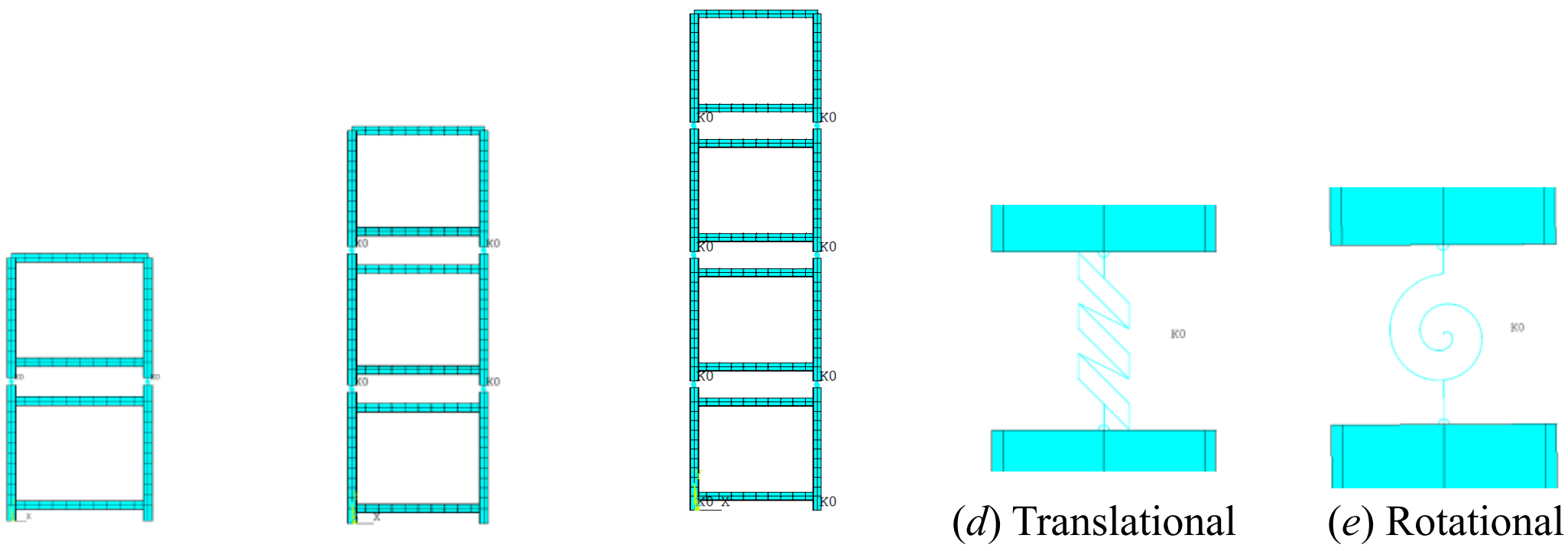

(*a*) 2S (*b*) 3S (c) 4S (*d*) Translational SD (*e*) Rotational SD

**Figure 20.** Model establishment using FEM (SD: spring-damper).

To determine the dynamic characteristics of the structure, conducting modal analysis is essential. In this study, the covariance-driven stochastic subspace identification method (SSI-COV) was employed for modal identification, chosen for its proven reliability and robust performance. Acceleration data were utilized to extract the first three bending modes, including natural frequencies, mode shapes, and damping ratios. The final results were averaged across 20 test sets.

$$\mathrm{MAC}(\mathbf{\Phi}_i, \mathbf{\Phi}_\mathrm{j}) = \frac{({\mathbf{\Phi}_i}^T \mathbf{\Phi}_j)^2}{({\mathbf{\Phi}_i}^T \mathbf{\Phi}_i)({\mathbf{\Phi}_j}^T \mathbf{\Phi}_j)} \tag{16}$$

To optimize structural parameters, model updating was performed using the first three modal frequencies as the target parameters to be adjusted. The translational and rotational spring stiffness at the base of the structure and each IMC required updating. The Artificial Ecosystem-Based Optimization (AEO) algorithm was employed for parameter refinement due to its exceptional convergence performance. A comprehensive comparison of the modal parameters of the numerical model before and after updating, in relation to the measured values, is presented in Table 9. The Modal Assurance Criterion (MAC) was utilized to evaluate the similarity of their mode shape vectors. A value approaching 1 indicated a higher degree of similarity between the mode shapes, which could be calculated using Equation (16).

**TABLE 9** First three natural frequencies and the MAC of modal shapes between measurement and FEM.

| 1st Mode | 2nd Mode |
|---|---|

| | Frequency(Hz) | | Damping ratio | MAC | Frequency(Hz) | | Damping ratio | MAC |
|---|---|---|---|---|---|---|---|---|
| | **Measuremet** | **FEM** | | | **Measuremet** | **FEM** | | |
| **2S** | 29.15 | 28.85 | 0.0049 | 0.90 | 40.07 | 39.99 | 0.0051 | 0.58 |
| **3S** | 16.74 | 16.68 | 0.0024 | 0.94 | 44.42 | 43.37 | 0.0015 | 0.73 |
| **4S** | 10.02 | 10.01 | 0.018 | 0.89 | 43.95 | 42.93 | 0.0036 | 0.72 |

| | 3th Mode | | | |
|---|---|---|---|---|
| | **Frequency(Hz)** | | **Damping ratio** | **MAC** |
| | **Measuremet** | **FEM** | | |
| **2S** | 85.01 | 78.36 | 0.0036 | 0.53 |
| **3S** | 76.47 | 88.99 | 0.0025 | 0.38 |
| **4S** | 77.86 | 72.63 | 0.0017 | 0.41 |